\documentclass[aps,prd,longbibliography,nofootinbib,amsthm,amsmath,amssymb,amsfonts,notitlepage]{revtex4-1}
\usepackage[utf8]{inputenc}
\usepackage[T1]{fontenc}
\usepackage[english]{babel}
\usepackage{graphicx}
\usepackage{float}
\usepackage{cleveref}
\usepackage{xcolor}

\newcommand{\bA}{{\bf A}}

\newcommand{\bb}{{\bf b}}
\newcommand{\bk}{{\bf k}}
\newcommand{\bP}{{\bf P}}
\newcommand{\bp}{{\bf p}}
\newcommand{\bQ}{{\bf Q}}
\newcommand{\br}{{\bf r}}

\newcommand{\bbe}{{\bf e}}

\newcommand{\lr}[1]{ \langle #1 \rangle}
\newcommand{\LG}{\mathrm{LG}}
\newcommand{\PW}{\mathrm{PW}}
\newcommand{\Minv}{M_{{\tiny \rm inv}}}
\newcommand{\keV}{\mathrm{keV}}
\newcommand{\nm}{\mathrm{nm}}

\newcommand{\doublet}[2]{ \left(\!\!\begin{array}{c}#1 \\ #2 \end{array}\!\!\right) }

\def\lsim{\mathrel{\rlap{\lower4pt\hbox{\hskip1pt$\sim$}}
		\raise1pt\hbox{$<$}}}         

\def\gsim{\mathrel{\rlap{\lower4pt\hbox{\hskip1pt$\sim$}}
		\raise1pt\hbox{$>$}}}         

\date{\today}

\begin{document}

\title{Universal features of high-energy scattering of Laguerre-Gaussian states}
\author{Yaoqi Yang}
\email{yangyq89@mail2.sysu.edu.cn}
\author{Igor P. Ivanov}
\email{ivanov@mail.sysu.edu.cn}
\affiliation{School of Physics and Astronomy, Sun Yat-sen University, 519082 Zhuhai, China}

\begin{abstract}
Vortex states of photons, electrons, and other particles are wave packets that carry  
intrinsic orbital angular momentum (OAM) and exhibit other features unavailable for plane waves.
Collisions of high-energy vortex states can become a promising tool for nuclear and particle physics, 
once experimental challenges are overcome.
An extensive literature exists on scattering processes involving vortex states; 
however, most works rely on assumptions that will be challenging to achieve in experiment.
In this work, we initiate a systematic re-analysis of vortex-state scattering processes
using paraxial Laguerre-Gaussian (LG) wave packets 
colliding at a non-zero impact parameter $b$. 
Since the total final transverse momentum $\bP_\perp$ is no longer fixed,
we focus on how the differential cross section depends on $\bP_\perp$.
We emphasize that non-trivial $\bP_\perp$-dependent features 
can originate either from the shape of the LG wave packets or from the dynamics of the scattering process under interest.
Here, we focus on the former source and explore in detail these universal kinematic features, 
while the study of process-specific modifications, along with the novel insights they may bring, is delegated to a future work.
Interestingly, the non-zero impact parameter $b$ plays a key role in many $\bP_\perp$-dependent effects,
making it a useful probe of vortex states, not a nuisance factor as often assumed.
\end{abstract}

\maketitle

\section{Introduction}

\subsection{Vortex state scattering as a novel tool in subatomic physics}

We explore the structure and dynamics of subatomic particles through their scattering.
Scattering is usually computed with the initial state particles chosen as plane waves (PW).
The fact that the real one-particle states are localized wave packets does not usually have a significant impact
on the predicted final state distributions.
In most quantum field theory (QFT) textbooks, the intermediate calculations often involve broad wave packets \cite{Peskin:1995ev}, 
and one usually takes the PW limit for the final results. 
This is justified in nearly all situations at colliders; although the processes with non-trivial final size effects exist, 
they are rare \cite{Kotkin:1992bj}.

About two decades ago, it was realized that the initial state photons, electrons, and other particles
can be prepared as the so-called {\em vortex}, or twisted, states. A vortex state is a wave packet with helical wave fronts 
that, on average, propagates along the axis $z$ and possesses the phase factor $\exp(i\ell\varphi_r)$ in the transverse plane, 
with an integer winding number $\ell$.
A single-particle state prepared as a vortex state carries an intrinsic adjustable $z$-projection of the orbital angular momentum (OAM).
For particles with spin, such as photons or electrons, exact vortex solutions of the Maxwell and Dirac equations have also been constructed,
see the recent review \cite{Ivanov:2022jzh} and references therein.

Vortex states have been generated in experiment.
The initial idea came from optics, where vortex photons and structured light have been used for decades \cite{Allen:1992zz};
see the recent reviews \cite{Knyazev:2019,forbes2021structured,Das:2026}.
In 2010, following the suggestion of Ref.~\cite{Bliokh:2007ec}, three research groups reported 
experimental production of vortex states of electrons of moderately relativistic energies, up to 300~keV \cite{Uchida:2010,Verbeeck:2010,McMorran:2011}.
Vortex electron research was quickly picked up by other teams, leading to the observation of peculiar effects
and the development of novel diagnostic tools, see reviews \cite{Bliokh:2017uvr,Lloyd:2017} and references therein.
Recently, composite particles such as cold neutrons \cite{Sarenac:2022} and slow atoms \cite{luski2021vortex} 
were also put in vortex states.

The adjustable OAM carried by vortex states represents a completely new degree of freedom, 
never used in nuclear and particle physics experiments.
Just as we measure the dependence of scattering cross sections on initial energy and polarization, we can also explore
how vortex-state cross sections depend on the OAM quantum number $\ell$. 
However, this is not the only novelty brought up by vortex states.
Scattering of vortex states, represented by specific PW superpositions, leads to new dimensions 
in the final phase space distribution that are unavailable for PW scattering \cite{Ivanov:2011kk}.
As a promising example, one can consider collisions of vortex electrons with vortex protons.
For elastic scattering, it will lead to novel form factors (correlations of amplitudes corresponding to different momentum transfers),
while for inelastic scattering it will help probe the nucleon excitations amplitudes
in a way complementary to the usual probes.
In the deep inelastic scattering regime, the presence of new initial state degrees of freedom
will force one to introduce new structure functions, and it is intriguing to see
how they will complement the traditional unpolarized and polarized structure functions of the plane wave 
deep inelastic scattering.
Moreover, one can prepare initial-state vortex wave packets in exotic spin-OAM entangled states \cite{Sarenac:2018evf,Li:2025pln}, 
which opens a new degree of freedom for spin physics, with tantalizing opportunities 
to probe the spin structure of the proton \cite{Deur:2018roz,Ji:2020ena} through alternative observables. 

To make vortex scattering a useful tool for nuclear and particle physics,
one needs to generate vortex particles with energies in the MeV and GeV ranges and bring them into collisions. 
Traditional particle sources and accelerators are incapable of shaping the wave front of individual particles
in a desired way; thus, new instrumentation is required.
Currently, the experimental efforts aimed to produce high-energy vortex photons, electrons, and other particles
include the generation of vortex soft X rays at the SOLEIL synchrotron
\cite{Carrara:2025adg}, the recent evidence for vortex 0.8~MeV gamma rays produced in an all-optical 
inverse Compton scattering experiment at Shanghai Jiao Tong University \cite{Wei:2025zsv},
the development of specialized linac injector equipment designed to produce ultrarelativistic vortex electrons \cite{Dyatlov:2025tyt},
and vortex-state-related activity at the China Advanced Nuclear Physics Research Facility in Huizhou \cite{An:2025lws}.
In addition, there are numerous proposals for the production of high-intensity vortex X rays at XFEL facilities \cite{Morgan:2022vuj,Zhou:2024zub},
for the acceleration of vortex electrons in linacs \cite{Meng:2025isy,Murtazin:2025gic,Karlovets:2025lnq},
and for relying on free-space collision process itself to upscatter vortex states to higher energies \cite{Jentschura:2010ap,Jentschura:2011ih,Ivanov:2011bv,Karlovets:2022evc,Karlovets:2022mhb}.
Similar ideas are also discussed for the production of vortex positrons \cite{Lei:2023wui,Ababekri:2024cyd}, protons, and other particles.

In short, vortex states enable novel scattering experiments in nuclear and particle physics, 
which will complement conventional scattering settings \cite{Ivanov:2022jzh,Ivanov:2026UJP}. 
Considering the ongoing and planned experimental R\&D on high-energy vortex states, it is timely to systematically compute 
differential cross sections of vortex scattering processes under realistic assumptions and highlight the key observables.

\subsection{A brief summary of vortex-state calculations}

There exists an extensive literature on the theoretical calculation of high-energy scattering processes involving vortex states 
\cite{Ivanov:2022jzh}, most of which deal with quantum electrodynamics (QED) processes.
Without attempting to review or mention all of them, we outline below typical questions addressed in these works
and highlight a gap that exists between theoretical investigation and future vortex scattering experiments.

Historically, the first example considered in \cite{Jentschura:2010ap,Jentschura:2011ih} was Compton scattering
of an optical vortex photon on a high-energy plane-wave electron. The key idea is that the final photon, scattered
strictly backward, takes a significant part of the electron's energy while preserving its vortex state.
The computations were done using monochromatic Bessel vortex beams, which are convenient for analytical calculations
but not normalizable in the transverse plane. If the final photon is scattered at a small, non-zero angle 
with respect to the initial electron direction, the vortex state is still approximately preserved \cite{Ivanov:2011bv}.
The recent indirect evidence of 0.8 MeV vortex photons reported in \cite{Wei:2025zsv} can be viewed as the first 
illustration of this technique. Alternatively, one can consider the scattering of a vortex high-energy electron
on a PW photon \cite{Seipt:2014bxa} and use the ring-shaped angular distribution of the final photon 
as a diagnostic tool for the vortex electron state.

It is important to recognize that the PW+V collision setting (a plane wave colliding with a vortex state)
cannot reveal the most interesting structures within the initial vortex state.
Specifically, the total PW+V cross section is given by the azimuthally averaged conventional PW+PW cross section
\cite{Ivanov:2011kk,Karlovets:2020odl}, up to an inessential flux modification factor.
Therefore, one should not expect a dramatic change in the cross section when replacing a single PW with a vortex state.
The event rate will change due to a different luminosity function, but not the total cross section.

As for the differential PW+V cross section, it does contain non-trivial effects due to a modified angular distribution, 
but these are not driven by the phase structure of the vortex state.
True interference-driven vortex effects in PW+V scattering occur only when one or more final state particles are projected onto vortex states.
This approach was taken in the first publications \cite{Jentschura:2010ap,Jentschura:2011ih,Ivanov:2011bv}
as well as in later works on linear and non-linear Compton scattering such as \cite{Taira:2016jbl,Chen:2018tkb}. 
The same kinematic framework was used for other QED processes, such as $\gamma\gamma \leftrightarrow e^+e^-$ \cite{Lei:2023wui,Ababekri:2024cyd,Liao:2026gqh}. 
It should be stressed, however, that projecting final-state particles onto vortex states must be done with care. 
Since the two final particles are momentum-entangled, 
they cannot be unambiguously assigned to well-defined vortex states.
Instead, the wave function of the entire final state must be considered,
and specific generalized measurement protocol that depends on detector details should be applied 
\cite{Karlovets:2022mhb,Karlovets:2022evc}.
Note that projecting the wave function onto a specific vortex state of an outgoing particle 
poses its own challenges; no currently available detector can perform such a projection for high-energy particles.

There are papers that do consider the collision of two vortex states, such as \cite{Ivanov:2012na,Ivanov:2016oue,Karlovets:2016dva,Karlovets:2016jrd,Korchagin:2024nen},
which predicted a novel quantum interference effect observable through interference fringes
in the total final momentum distribution. 
The showcase example considered in \cite{Ivanov:2012na,Ivanov:2016oue} was M\o{}ller (elastic electron-electron) scattering;
similar effects were recently studied for vortex $p\bar p \to e^+e^-$ \cite{Korchagin:2024nen}. 
In these papers, the vortex states were treated as Bessel beams, 
with additional smearing of the Bessel-state parameters to render the expressions finite.
Although this procedure is technically correct, it introduces arbitrary parameters, which reduces the predictive power of the results.
Another recent paper \cite{Zhao:2023cwd} examined kinematic features 
that emerge in the scattering of two Laguerre-Gaussian wave packets. However, that work had limitations:
it did not address specific processes, the calculations were performed numerically, 
and the dependence on the impact parameter was not discussed.

Although the above works set the stage for future research and provide interesting insights, 
their calculations often rely on assumptions about vortex state parameters that will be extremely difficult to achieve in future experiments.
Some of them use a very large vortex cone opening angle $\theta \sim 1$, 
a key parameter that characterizes the tight focusing of a vortex state. 
It is unlikely that such states can be realized in the near future, 
as all experimental schemes proposed so far for producing high-energy vortex states lead to small opening angles.
Additionally, some authors assume that the two vortex states are defined with respect to the same vortex axis.
In real situations, one should expect a non-zero impact parameter $b$ between the axes of the two vortex beams.

In summary, the landscape of theoretical exploration of vortex particle scattering and predictions of testable effects appears rather eclectic.

\subsection{Goals of the present work}

In this work, we initiate a project aiming to revise vortex scattering calculations 
within the formalism of paraxial Laguerre-Gaussian (LG) states
based on a set of assumptions that we believe will be close to future experiments. 

The key observables in LG-LG wave packet collisions are defined in terms of the total final state momentum,
which are not available for PW scattering.
These features can arise from two distinct sources: the vortex nature of the initial-state LG wave packets 
and the properties of the scattering process itself.
The former is ``instrumental'' as it reflects how we prepare the vortex states,
while the latter is related to the scattering amplitude and may reveal non-trivial interaction dynamics.
We are usually interested in the latter source.
However, to accurately interpret any peculiarities in the cross section that may emerge in future experiments, 
we must clearly disentangle these two sources.

This is why we decided to delegate the analysis of specific QED processes to follow-up works and 
focus in this paper on the systematic exploration of the first source of non-trivial effects.
We will examine the universal kinematic features of the differential cross section of 
a generic non-resonant $2\to 2$ LG-LG scattering process at non-zero impact parameter. 
We will describe several remarkable effects, some of which have been previously reported for specific processes.
We will study how these effects depend on the initial OAM and the impact parameter.
The results of this work will serve as a reference point, upon which future discussions 
of additional process-dependent effects can be built.

Although we aim to explore universal kinematic features, for the reader's convenience, we will frame our discussion 
in the context of M\o{}ller scattering of two vortex electrons.
However, we stress that a detailed study of the vortex M\o{}ller scattering cross section 
and other specific QED processes will be presented in follow-up papers.

\subsection{Assumptions adopted}

To clarify how our approach differs from the previous studies, let us first list our assumptions.
We consider the free-space collision of two spatially localized LG wave packets. We do not consider fixed target collisions, 
such as the scattering of an electron wave packet on a trapped atom, nor the intersection of infinitely long monochromatic beams. 
We adopt the paraxial approximation for the LG wave packets, which is applicable to nearly all vortex states experimentally produced to date.
We also neglect the spreading of the wave packets during the collision event (the impulse approximation). 
When dealing with massive particles, we derive results using general relativistic kinematics,
without taking non-relativistic or ultra-relativistic limits.

The final state particles are described by plane waves, just as in the usual scattering calculations.
We assume that the final particles are detected with conventional pixelized detectors with sufficient accuracy. 
Our key observable will be the dependence of the differential cross section on the total transverse momentum.

We will consider both zero and non-zero impact parameters. 
Contrary to common wisdom, which views a non-zero impact parameter as an undesirable effect,
we find that it plays an instrumental role in revealing several vortex-induced effects. 

Since we focus on generic, process-independent kinematic effects, we do not need to include the spin degrees of freedom.
These will be addressed in future works when we extend our analysis to specific processes.

The structure of this paper is as follows. In section~\ref{section-vortex}, 
we remind the reader of the formalism for wave packet scattering calculations, 
together with the simplifications resulting from the paraxial and impulse approximations.
We present a general expression for a $2\to2$ scattering cross section
as the plane-wave cross section times a new factor that exhibits the distribution in the total final momentum $\bP$.
This factor, especially its transverse momentum dependence, is the key quantity that reveals the characteristic
vortex scattering effects.
Then, in section~\ref{section-universal}, we focus on the universal, process-independent features of this factor.
We begin by evaluating the transverse momentum integral analytically, for both 
zero and non-zero impact parameters, discuss the intriguing effects that emerge.
Next, in section~\ref{section-insights} we present a more detailed exploration of the three effects
that arise at a non-zero impact parameter:
a peculiar transverse momentum imbalance along with the superkick effect, 
high-contrast interference fringes that exist only for rather specific arrangements,
and the previously unknown effect of a controllable splitting, in momentum space, 
of a single vortex into two vortices.
We conclude with a discussion and a summary of our findings.
Intermediate technical computations, including an exposition of the method for calculating LG overlap integrals
at a non-zero impact parameter, are presented in the appendix.

Throughout the paper, we use the relativistic units $\hbar = 1$ and $c = 1$ 
and denote the 3-vectors with bold symbols, adding the subscript $\perp$ for transverse vectors.

\section{Scattering of two vortex states: an overview of the formalism} \label{section-vortex}

\subsection{General expressions}

To remind the reader of the general framework used for calculating scattering of arbitrarily shaped wave packets,
let us begin with the textbook case of plane-wave scattering. 
We consider, for definiteness, our showcase example of elastic electron-electron scattering;
the formalism can be readily extended to any other $2\to2$ processes. 
The two initial particles are plane waves with three-momenta $\bk_1$, $\bk_2$ and energies $E_1$, $E_2$.
The two final particles are described with three-momenta $\bk'_1$, $\bk'_2$ and energies $E'_1$, $E'_2$;
their total momentum and energy are $\bP = \bk'_1 + \bk'_2$ and $E_f = E_1' + E_2'$. 
The plane-wave $S$-matrix element has the form
\begin{equation}
S_{\PW}(k_1,k_2;k'_1,k'_2) = i(2\pi)^4\delta^{(4)}(k_1+k_2 - k_1'-k_2') {{\cal M} \over \sqrt{16 E_1 E_2 E'_1 E'_2}}\,, 
\label{SPW}
\end{equation}
where we tacitly assume the usual PW normalization of unit flux.
Here, the invariant amplitude ${\cal M}$ depends on all four momenta, ${\cal M}(\bk_1,\bk_2;\bk_1',\bk_2')$. 
The plane wave scattering cross section can then be written as
\begin{equation}
d\sigma_{\PW} = \frac{(2\pi)^4\delta^{(4)}(k_1+k_2 - k_1'-k_2') |{\cal M}|^2}{4I_{\PW}}\cdot \frac{d^3k_1'}{(2\pi)^3 2E_1'}\, 
\frac{d^3k_2'}{(2\pi)^3 2E_2'}\,.\label{PW-cross-section}
\end{equation}
where $I_{\PW} = \sqrt{(k_1 \cdot k_2)^2-m^4}$ is the flux invariant.

Clearly, for fixed initial $\bk_1$ and $\bk_2$, the total final state momentum is also fixed: $\bk_1 + \bk_2 = \bP = \bk_1' + \bk_2'$. 
Working in the center of mass frame and performing the necessary final phase space integration, 
we obtain the well-known angular distribution 
\begin{equation}
d\sigma_{\PW} = \frac{|{\cal M}|^2}{64\pi^2 E_0^2}d\Omega_1 = \frac{|{\cal M}|^2}{64\pi^2 E_0^2} \frac{d^2 k_{1\perp}'}{|\bk_{1}'|\, k_{1z}'}\,.\label{dsigma-PW}
\end{equation}
Here, $\Omega_1$ refers to the solid angle of the first final particle; 
the momentum of the second final particle is $\bk_2' = -\bk_1'$.
The latter form of the cross section is especially useful for the small-angle scattering, when $|\bk_{1\perp}'| \ll |\bk_{1}'|$ and $k_{1z}' > 0$.
Although we do not explicitly write the spin degrees of freedom for the initial and final particles,
it is understood that ${\cal M}$ represents the plane-wave helicity amplitude for any chosen set of initial and final helicities.
If needed, the unpolarized cross section can then be computed in the standard way by averaging the cross section 
over the initial helicities.

Suppose now that the two initial particles are prepared as localized wave packets. 
Let us disregard their spin degrees of freedom.
The scattering theory of arbitrarily shaped, partially coherent beams
was first developed in \cite{Kotkin:1992bj} within the paraxial approximation,
which is what we adopt in this work; the extension of this formalism 
beyond the paraxial approximation can be found in \cite{Karlovets:2016jrd,Karlovets:2020odl}.
We describe the two initial particles as momentum space wave packets $\phi_1(\bk_1)$ and $\phi_2(\bk_2)$ normalized according
to the Lorentz-invariant prescription of \cite{Karlovets:2020odl,Karlovets:2018iww}:
\begin{equation}
\int \frac{d^3 k}{(2\pi)^3}\, \frac{1}{2E(k)}\, |\phi_i(\bk)|^2 = 1\,,
\end{equation}
with $E(k) = \sqrt{\bk^2+m^2}$.
We work in generic relativistic kinematics and do not adopt the non-relativistic or ultrarelativistic approximation.
The average momenta of the two colliding wave packets are $\bp_1 = \langle \bk_1\rangle$ and $\bp_2 = \langle \bk_2\rangle$. 
We assume that these average momenta are antiparallel and define the common axis $z$,
with $p_{1z} > 0$, $p_{2z} < 0$;
at present, we do not require them to sum up to zero: $p_{1z} \not = |p_{2z}|$.
The average energy of each wave packet can be defined either as $\varepsilon_i = \sqrt{p_i^2 + m^2}$ or $\lr{E_i(k_i)}$.
Since their difference is negligible within the paraxial approximation, we use $\varepsilon_i$. 
Nevertheless, it is important to keep in mind that the energy of individual PW components in the wave packet are not fixed:
$E_i(k_i) \neq \epsilon_i$. 
Finally, whenever we need the coordinate wave functions, we define them via
\begin{equation}
	\psi(\br,t) = \int \frac{d^3k}{(2\pi)^3\sqrt{2E(k)}}\,\phi(\bk)\, e^{i\bk\br - iE(k)t},\quad
	\int d^3r\, |\psi(\br,t)|^2 = 1\,.\label{psi-def}
\end{equation}

We now turn to the vortex states and specify the momentum space wave functions following \cite{Karlovets:2020odl}.
Both initial-state particles are described as relativistic Laguerre-Gaussian (LG) principal mode wave packets:
\begin{equation}
	\LG(\bk \,|\, \bp, \ell, \sigma_{\perp}, \sigma_{z}) = \sqrt{2E}\, (4\pi)^{3/4}\sigma_{\perp}\sqrt{\sigma_{z}} 
	\frac{(\sigma_{\perp}k_{\perp})^{|\ell|}}{\sqrt{|\ell|!}}	\cdot 
	\exp\left[ -\frac{k_{\perp}^2 \sigma_{\perp}^2}{2}-\frac{(k_{z}-p_{z})^2 \sigma_{z}^2}{2} +i\ell\varphi_{k}\right] \,.
	\label{LG-general}
\end{equation}
The integer $\ell$ represents the winding number of the LG state. We allow $\ell$ to be of either sign.
In the expression for $\LG(\bk \,|\, \bp, \ell, \sigma_{\perp}, \sigma_{z})$, 
the momentum $\bk$ is the argument while the remaining labels are parameters.
Here, $\sigma_{\perp}$ and $\sigma_{z}$ are the transverse and longitudinal spatial extents and are taken as independent parameters. 
In particular, we choose $\sigma_{z}$ to be of the same order as $\sigma_{\perp}$
in order to guarantee that the wave packets do not significantly spread during the overlap time, so that 
the all-important impulse approximation is applicable \cite{Ivanov:2022sco}. 

Note that we do not claim that the expression \eqref{LG-general} is Lorentz invariant, 
nor even form-invariant under large Lorentz boosts.
The shape of the LG wave packet primarily depends on the way it is produced and propagated to the collision point.
Thus, Eq.~\eqref{LG-general} is an Ansatz, which implicitly relies on working in the laboratory frame.
We are ready to accept the potential limitations of our calculations. We will always work in the laboratory frame
and are not going to switch, for example, to the rest frame of one of the colliding particles.
For completeness, we mention that a Lorentz invariant formulation of the LG wave packets, both paraxial and non-paraxial, 
was developed in \cite{Karlovets:2018iww}.
We also mention in passing that one can also go beyond the principal modes and employ the LG modes 
with the radial quantum number $n > 0$ as well as the so-called elegant LG wave packets \cite{Sheremet:2024jky}.
Investigating these modifications is beyond the scope of the present paper.

In our collision process, we take into account a non-zero transverse impact parameter $\bb_\perp$ between the vortex axes 
of the two initial particles:
\begin{equation}
	\phi_1(\bk_1) = \LG(\bk_1\,|\,  \bp_1, \ell_1, \sigma_{1\perp}, \sigma_{1z})\,, \quad
	\phi_2(\bk_2) = \LG(\bk_2\,|\,  \bp_2, \ell_2, \sigma_{2\perp}, \sigma_{2z})
	\cdot e^{-i \bb_\perp \bk_{2\perp}}\,.
	\label{LG-1-2}
\end{equation}
In this way, we assume that the focal spots of the two LG wave packets are located at the same $z$ but at different transverse coordinates;
moreover, we also assume that the instants of maximal focusing of the two wave packets coincide.
This is, of course, not the most general arrangement. In principle, we could have introduced additional parameters 
that would allow for a longitudinal and temporal mismatch of the two focusing events. 
However, as demonstrated in \cite{Liu:2022nfq}, these become redundant parameters under the impulse approximation.
We admit that collisions of longitudinally extended beams would require introduction such shifts and would necessitate
a rigorous treatment beyond the impulse approximation. However, we believe that in all realistic high-energy scattering processes
this approximation is justified.

We stress that only the initial particles are taken as LG wave packets. 
For the final particles, we use the standard PW basis.

With all these definitions, we present the generalized cross section as in \cite{Kotkin:1992bj,Karlovets:2020odl}:
\begin{equation}
d\sigma(\bk_1',\bk_2') = \frac{1}{L}\cdot(2\pi)^8 |{\cal I}|^2 \frac{d^3k_1'}{(2\pi)^3 2E_1'}\, \frac{d^3k_2'}{(2\pi)^3 2E_2'}\,.
\label{dsigma-LGLG}
\end{equation}
Here, instead of the PW scattering amplitude ${\cal M}$, we deal with the wave packet scattering amplitude
\begin{equation}
{\cal I} = \int\frac{d^3k_1}{(2\pi)^3 2E_1} \frac{d^3k_2}{(2\pi)^3 2E_2}\,\phi_1(\bk_1)\,\phi_2(\bk_2)\,
\delta^{(4)}(k_1+k_2 - k_1'-k_2')\cdot {\cal M}\,.\label{cal-I}
\end{equation}
The quantity $L$ represents the luminosity function for collision of two wave packets \cite{Kotkin:1992bj,Karlovets:2020odl}. 
In the exact, non-paraxial treatment, $L$ can be expressed in terms of Wigner's functions for the two colliding wave packets \cite{Karlovets:2020odl}.
However, within the paraxial approximation, the relative velocity $|v_1-v_2|$ can be computed 
via the average momenta of the two wave packets $v_i = p_{iz}/\varepsilon_i$ (with $v_1>0$ and $v_2<0$).
Then it can be taken out of the integral, leading to the luminosity function in the form of the space-time overlap of the two colliding wave packets:
\begin{equation}
	L = |v_1 - v_2| \int d^3 r\, dt\, |\psi_1(\br,t)|^2 \, |\psi_2(\br,t)|^2\,.\label{lumi}
\end{equation}
This factor takes care of the correct normalization of the cross section. 
This is especially important when the two colliding wave packets overlap only partially,
as is the case for a significant impact parameter $\bb_\perp$.
Indeed, in the case of a near ``miss'', the scattering probability is strongly suppressed
but so is the luminosity function $L$. As a result, the cross section $d\sigma$ does not experience any significant reduction.
This is, of course, very well known in collider physics; when the two beams are separated as in van der Meer scans,
the event rate drops sharply but its decrease is never attributed to the cross section modification.

Let us also briefly discuss the role of the polarization degrees of freedom. 
If the PW amplitude ${\cal M}$ is understood as a helicity amplitude, the initial-state spinors (for fermions)
or polarization vectors (for photons or other vector fields) is not constant but depends on the momentum $\bk_i$. 
There exist several ways the vortex fermion and vortex photon could be constructed,
see reviews \cite{Bliokh:2017uvr,Knyazev:2019,Ivanov:2022jzh} for various options.
The most natural and rigorous way is to construct exact monochromatic solutions of the Dirac or Maxwell's equations
in cylindrical coordinates and built the Bessel fermion and photon states with definite energy, longitudinal momentum,
the modulus of the transverse momentum, the total angular momentum $z$-projection $m$ (half-integer for fermions), 
and a definite helicity. Then, one could further combine these solutions into localized wave packets such as Bessel-Gaussian states.
Note that the OAM $\ell$ and the spin $z$-projections $s_z$ are not, strictly speaking, conserved even in free space.
However, as known already from the early days of vortex photons \cite{Allen:1992zz}, in the paraxial approximation,
this spin-OAM interaction can be neglected and one can speak of approximately conserved OAM and spin.
This is the pragmatic strategy we implicitly use in the present paper; 
its realization for specific QED processes will be presented elsewhere.

\subsection{Increased dimensionality of the final state phase space}

\begin{figure}[!h]
	\centering
	\includegraphics[width=0.7\textwidth]{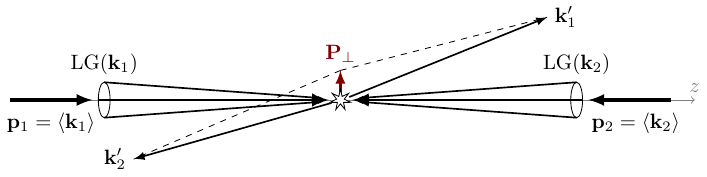}
	\caption{Schematic layout of average center of mass scattering of two initial LG states into two final plane waves with momenta $\bk_1'$ and $\bk_2'$.
	The transverse part of the total final state momentum $\bk_{1\perp}' + \bk_{2\perp}' = \bP_\perp$ is non-zero and lies in 
	a region of the order of $1/\sigma_{i\perp}$.}
	\label{fig-scattering-geometry}
\end{figure}

Since the four-dimensional delta-function that encodes the energy-momentum conservation is absorbed
in the definition of the integral ${\cal I}$ in Eq.~\eqref{cal-I}, 
it no longer appears in the differential cross section given in Eq.~\eqref{dsigma-LGLG}.
This implies that the total final momentum $\bP$, the total final energy $E_f$, and the final system invariant mass
defined via $\Minv^2 = E_f^2 - \bP^2$ are no longer fixed but are distributed within certain ranges.
Writing $d^3k_1'd^3k_2' = d^3k_1'd^3P$, we can now fix $\bk_1'$ and study the distribution in $\bP$.
This is the new dimension in the final phase space that opens up in vortex state collisions.

As first remarked in \cite{Ivanov:2011kk}, the differential cross section exhibits in this dimension new features (interference fringes) 
that originate from interference between different pairs of PW components inside $\phi_1(\bk_1)$ and $\phi_2(\bk_2)$.
Not only does this interference represent a new phenomenon worth verifying in experiment, 
but it also helps reveal new details of scattering dynamics, which the PW collisions are blind to. 
The two prominent examples are the sensitivity of the collision process to the overall phase of the scattering amplitude
\cite{Ivanov:2012na,Ivanov:2016oue,Karlovets:2016dva} and the intuition-defying phenomenon dubbed the superkick effect
\cite{Li:2024mzd,Liu:2025jei} that appears already for LG-Gaussian collision.
These features make vortex state collisions a valuable particle dynamics probe, complementary to the traditional particle collision settings.

To further highlight the importance of the $\bP$-space distribution, let us demonstrate that integrating the cross section over the $\bP$-space
removes most of the novel features of the vortex state collisions and brings us back to the good old plane-wave result.
First, following \cite{Liu:2022nfq}, we express the final phase space measure in \eqref{dsigma-LGLG} as 
\begin{equation}
	\frac{d^3k_1'd^3k_2'}{E_1'E_2'} = \beta(\Minv)\frac{1}{2} d\Omega_{1,{\tiny\rm cm}} d\Minv d^3P\,.
\end{equation}
The invariant mass depends on $\bP$. Since $\bP$ is not fixed, we do not have a unique center of mass frame of the two final particles.
Instead, we must deal with individual center of mass frames for each value of $\bP$. Still, in every such frame, 
we can define the velocity of each final particle $\beta(\Minv) = \sqrt{1-4m^2/\Minv^2}$ 
and the solid angle element of the first final particle $d\Omega_{1,{\tiny\rm cm}}$. 
This procedure is analogous to the standard description of the phase space of three-particle systems.

Next, we now assume that the initial wave packet momenta are chosen so that $\bp_1 + \bp_2 = 0$,
that is, the laboratory frame coincides with the wave packet center of mass frame.
Let us furthermore assume\footnote{We do not claim that this approximation is warranted for all processes and in all kinematic situation.
It does not apply for production of narrow resonances or for M\o{}ller scattering near the forward peak.
We use it here only to illustrate that in a generic situation, the full integration over the $\bP$ space
washes out interesting effects.} that the plane-wave scattering amplitude ${\cal M}$ 
is a smooth function of all the momenta and can be approximated by
\begin{equation}
{\cal M} \approx {\cal M}_0 = {\cal M}(\bp_1,\bp_2;\bk_1',\bk_2').\label{M0} 
\end{equation}
Put simply, we consider processes without narrow resonances or kinematic singularities;
elastic $ee$ scattering away from the forward region is of this type.
Then, replacing ${\cal M}$ with ${\cal M}_0$ allows us to take it out of the integral in Eq.~\eqref{cal-I},
which amounts to replacing ${\cal I}$ with
\begin{equation}
	{\cal I}_0 = {\cal M}_0\cdot \int\frac{d^3k_1}{(2\pi)^3 2E_1} \frac{d^3k_2}{(2\pi)^3 2E_2}\,\phi_1(\bk_1)\,\phi_2(\bk_2)
	\delta^{(4)}(k_1+k_2 - k_1'-k_2')\,.\label{cal-I0}
\end{equation}
At the same time, the difference between $\Minv$ and $E_f$ becomes negligible,
$\beta d\Omega_1$ becomes $\bP$-independent, and $d\Minv d^3P$ can be replaced with $d^4P$.
The differential cross section in Eq.~\eqref{dsigma-LGLG} can then be approximated as
\begin{equation}
	d\sigma = \frac{\pi^2}{2L} \beta d\Omega_1 \cdot |{\cal I}_0|^2 d^4P\,.
	\label{dsigma-LGLG-2}
\end{equation}
Now comes the main point. Suppose we do not wish to explore the fine structure of the cross section in the $\bP$ space 
and simply integrate it over all $d^4P$.
As shown in \cite{Liu:2022nfq}, this can be done without dealing with explicit wave functions: just use the definition of ${\cal I}_0$ 
in Eq.~\eqref{cal-I} and transform the four-dimensional delta function as
$(2\pi)^4\delta^{(4)}(k_1+k_2-P) = \int d^4x\, \exp[i(k_1 + k_2 - P)x]$.
Then, using this representation twice in $|{\cal I}_0|^2$, perform all the $k_i$ integrations and 
express the result as
\begin{equation}
	\int |{\cal I}_0|^2 d^4P = \frac{|{\cal M}_0|^2}{(2\pi)^4\, E_0^2} \int d^4x\, |\psi_1(\br,t)|^2 |\psi_2(\br,t)|^2\,.
	\label{I2d4P}
\end{equation}
The remaining integral is exactly the luminosity function \eqref{lumi} without the relative velocity, $L/|v_1-v_2|$.
Inserting all these result in \eqref{dsigma-LGLG-2}, we arrive at the PW differential cross section Eq.~\eqref{dsigma-PW}.

The take-home message is that the key kinematic novelties of generic non-resonant vortex scattering away from kinematic singularities
reside in the $\bP$ space.
By integrating over these new dimensions, one loses access to most of these novelties.

We remark however that this conclusion may become invalid for processes mediated by narrow resonances or in the vicinity
of kinematic singularities, such as M\o{}ller scattering near the forward region.
Such process-specific modifications will require special discussions delegated to future work.

\subsection{Transverse momentum differential cross section}

Localized wave packets evolve in time.
For LG states, it implies that the localization parameters $\sigma_\perp$ and $\sigma_z$
depend on time.
The exact expressions for the time evolution of the LG wave functions
were presented in \cite{Karlovets:2018iww,Karlovets:2020odl,Liu:2022nfq} and are summarized in Appendix~\ref{appendix-impulse}.
We believe that in all realistic situations of high-energy vortex particle collisions
this spreading can be neglected over the timescale of the event duration. 
Following \cite{Liu:2022nfq}, we call this assumption the {\em impulse approximation} and repeat
in appendix~\ref{appendix-impulse} the key steps of its derivation.
The net result is that we can replace the time-dependent 
$\sigma_\perp(t), \sigma_z(t)$ with their values at the moment of collision,
which leads to a considerable simplification of the wave packet scattering amplitude \cite{Liu:2022nfq}. 
As outlined in appendix~\ref{appendix-impulse}, the expression for ${\cal I}$ can be factorized
into the longitudinal and transverse parts:
\begin{equation}
	{\cal I} =  \frac{1}{(2\pi)^4\, \sqrt{\varepsilon_1\varepsilon_2|v_1-v_2|}}
	\frac{\sigma_{1\perp}\sigma_{2\perp}}{\sqrt{|\ell_1|!\,|\ell_2|!}}  
	\cdot {\cal I}_L \cdot {\cal I}_\perp\,.\label{cl-I-3}
\end{equation}
with ${\cal I}_L$ defined so that $\int |{\cal I}_L|^2 dP_z dE_f = 1$. It is the transverse part 
\begin{eqnarray}
	{\cal I}_\perp &=& 
	\int d^2 k_{1\perp}\,d^2 k_{2\perp}\, \delta^{(2)}(\bk_{1\perp} + \bk_{2\perp} - \bP_\perp)\, 
	(\sigma_{1\perp}k_{1\perp})^{|\ell_1|} \, (\sigma_{2\perp}k_{2\perp})^{|\ell_2|}  \, e^{i(\ell_1\varphi_1 + \ell_2\varphi_2)}\nonumber\\
	&&\quad \times \ \exp\left[ -\frac{k_{1\perp}^2 \sigma_{1\perp}^2}{2} -\frac{k_{2\perp}^2 \sigma_{2\perp}^2}{2} 
	- i\bb_\perp \bk_{2\perp}\right]	\cdot {\cal M}\,.\label{I-perp}
\end{eqnarray}
that is of particular interest, as it contains the interference patterns
in the total transverse momentum $\bP_\perp$ space.
In principle, one could follow the standard path: square ${\cal I}$, compute explicitly the luminosity function $L$, 
insert both expressions in the center of mass differential cross section in Eq.~\eqref{dsigma-LGLG-2}, 
and perform the integration in $P_z$ and $E_f$ to eliminate $|{\cal I}_L|^2$.
However, there exists a shortcut to the final expression.
Let us invert Eq.~\eqref{I2d4P} to express the luminosity function in the center of mass frame, in which $|v_1-v_2| = 2\beta$, in terms of ${\cal I}_0$:
\begin{equation}
	L = \frac{2\beta \cdot (2\pi)^4 E_0^2 }{|{\cal M}_0|^2} \int |{\cal I}_0|^2 d^4P\,.
\end{equation}
Substituting this expression directly into the differential cross section, we obtain
\begin{equation}
	d\sigma = \frac{|{\cal M}_0|^2}{64\pi^2 E_0^2} \,d\Omega_1 \, \frac{|{\cal I}|^2\, d^4P}{\int |{\cal I}_0|^2\, d^4P}\,.\label{dsigma-LGLG-3}
\end{equation}
Next, since the plane-wave scattering amplitude ${\cal M}$ appears only inside ${\cal I}_\perp$, the longitudinal
parts are the same in the numerator and denominator: ${\cal I}_L = {\cal I}_{0L}$.
Therefore, by integrating over $P_z$ and $E_f$, we obtain our final expression for 
the transverse momentum distribution:
\begin{equation}
	d\sigma(\bk_{1\perp}',\bP_\perp) 
	= d\sigma_{\PW}(\bk_{1\perp}') \cdot W d^2P_\perp\,,
	\quad \mbox{where}\quad W = \frac{|{\cal I}_\perp|^2}{\int |{\cal I}_{0\perp}|^2\, d^2P_\perp}\,.
	\label{W}
\end{equation}
The function $W(\bk_{1\perp}',\bP_\perp)$ shows the $\bP_\perp$ distribution of the differential cross section at various values of $\bk_{1\perp}'$,
and it will be our main focus in the following sections.\footnote{The asymmetric definition used here is a matter of choice.
If needed, the differential cross section \eqref{W} can be viewed as a function of $(\bk_{1\perp}',\bk_{2\perp}')$,
or $(\bk_{2\perp}',\bP_\perp)$; what is important is that we can study the transverse momentum distribution
of the first and, simultaneously, the second final particle.}

In general, the $\bP_\perp$ density $W$ defined in Eq.~\eqref{W} depends on the PW scattering amplitude ${\cal M}$ of the process we consider.
However, if the kinematics is such that ${\cal M}$ is a sufficiently smooth function
of the initial-state momenta, we can approximate it with ${\cal M}_0$, see Eq.~\eqref{cal-I0}.
Then, $W$ simplifies to
\begin{equation}
	W_0 = \frac{|{\cal I}_{0\perp}|^2}{\int |{\cal I}_{0\perp}|^2\, d^2P_\perp}\,,
	\quad \mbox{so that}\quad \int W_0\, d^2P_\perp = 1\,.
	\label{function-W0}
\end{equation}
This quantity is universal in the sense that it depends only on the configuration of the colliding wave packets 
but not on the scattering process itself.
This is why we need first to explore $W_0$ in the $\bP_\perp$-plane for various LG wave packet parameters 
and later, in follow-up papers, check which additional, process-specific features $W$ contains.

\section{Universal kinematic properties of LG-LG scattering}\label{section-universal}

\subsection{General properties of the transverse integral}\label{section-general-properties}

We now turn to the main part of the present work: the explicit evaluation of the transverse integral ${\cal I}_{0\perp}$
as a function of $\bP_\perp$ and $\bb_\perp$
and discussion of physical implications of its non-trivial dependence.
For future convenience, we now set the dummy quantity ${\cal M}_0 = 1$ and redefine 
the transverse integral as
\begin{eqnarray}
	{\cal I}_{0\perp} &=& 
	\int d^2 k_{1\perp}\,d^2 k_{2\perp}\, \delta^{(2)}(\bk_{1\perp} + \bk_{2\perp} - \bP_\perp)\, 
	(\sigma_{1\perp}k_{1\perp})^{|\ell_1|} \, (\sigma_{2\perp}k_{2\perp})^{|\ell_2|}  \, e^{i(\ell_1\varphi_1 + \ell_2\varphi_2)}\nonumber\\
	&&\quad \times \ \exp\left[ -\frac{k_{1\perp}^2 \sigma_{1\perp}^2}{2} -\frac{k_{2\perp}^2 \sigma_{2\perp}^2}{2} 
	- i\bb_\perp \cdot \bk_{2\perp}\right]	\,.\label{I0-redef}
\end{eqnarray}
As detailed in Appendix~\ref{appendix-Iperp}, one can switch to the coordinate-space integration over $\br_{i\perp}$ 
(shortened to $\br_i$) and represent the same integral as
\begin{equation}
	{\cal I}_{0\perp} =
	\frac{i^{|\ell_1|+|\ell_2|}}{\sigma_{1\perp}^{|\ell_1|+2} \sigma_{2\perp}^{|\ell_2|+2}} 
	\int d^2 r_{1}d^2 r_{2}\ \delta^{(2)}(\br_{1}-\br_{2}-\bb_\perp)\, 
	r_{1}^{|\ell_1|}\,e^{i\ell_1\varphi_{r1}} \, r_{2}^{|\ell_2|}\,e^{i\ell_2\varphi_{r2}}
	\exp\left[-\frac{r_{1}^2}{2\sigma^2_{1\perp}}-\frac{r_{2}^2}{2\sigma^2_{2\perp}}-i\bP_\perp \cdot \br_{1}\right].\label{I0-redef-r}
\end{equation}
The comparison of the two expressions reveals their close similarity, which implies that 
${\cal I}_{0\perp}$ as a function of $\bP_\perp$ at fixed $\bb_\perp$
should be structurally similar, as its dependence on $\bb_\perp$ at fixed $\bP_\perp$.
This is no surprise as it represents the built-in coordinate-momentum duality of the paraxial LG states, 
compare Eqs.~\eqref{LG-general} and~\eqref{LG-general-r}. 

Before proceeding to the evaluation of the transverse integral, we remark that essentially the same quantity 
$W_0$ was explored in \cite{Zhao:2023cwd} but only at zero impact parameter.
That study revealed distinct oscillatory patterns in the $\bP_\perp$-space 
for $\ell_1, \ell_2$ being of the same or opposite signs,
and drew an interesting comparison between the Bessel beam and LG wave packet collisions. 
However, the LG-LG collision case was treated in \cite{Zhao:2023cwd} only numerically, 
and the origin of the sharp contrast between
$\ell_1 \ell_2 > 0$ and $\ell_1 \ell_2 < 0$ patterns remained unclear.
Below, we give analytical expressions for the integral and study its dependence
on the impact parameter $\bb_\perp$, which plays a crucial role in several physics effects.

\subsection{Transverse momentum distribution: analytical results for $b = 0$}

For the zero impact parameter case, the calculations presented in Appendix~\ref{appendix-Iperp} give
\begin{equation}
	{\cal I}_{0\perp} \propto e^{i(\ell_1+\ell_2)\varphi_P}\, P_\perp^{\,|\ell_1+\ell_2|} 
	\exp\left(-\frac{P_\perp^2 \Sigma^2_{\perp}}{2}\right) \cdot L^{|\ell_1+\ell_2|}_{\ell_-}\left(\frac{P_\perp^2 \Sigma^2_{\perp}}{2}\right)\,,
	\quad \mbox{where}\quad \Sigma^2_{\perp} \equiv \frac{\sigma^2_{1\perp}\sigma^2_{2\perp}}{\sigma^2_{1\perp} + \sigma^2_{2\perp}}\,.
	\label{cal-I0perp}
\end{equation}
Here, $L^{|\ell_1+\ell_2|}_{\ell_-}$ denotes the associated Laguerre polynomial with the lower index given by
\begin{equation}
	\ell_- \equiv \frac{|\ell_1| + |\ell_2| - |\ell_1 + \ell_2|}{2} = 
	\left\{
	\begin{array}{l}
		0 \quad \mbox{for $\ell_1, \ell_2$ of the same signs,} \\[2mm]
		\min(|\ell_1|, |\ell_2|) \quad \mbox{for $\ell_1, \ell_2$ of the opposite signs.} 
	\end{array}	
	\right.
	\label{ell-minus}
\end{equation}
Although the prefactor for ${\cal I}_{0\perp}$ was also computed in Appendix~\ref{appendix-Iperp}, 
it is not essential here because it cancels in $W_0$: 
\begin{equation}
	W_0 = C_0\cdot (P^2_\perp \Sigma^2_{\perp})^{|\ell_1+\ell_2|} \,
	e^{-P_\perp^2 \Sigma^2_{\perp}} \cdot \left[L^{|\ell_1+\ell_2|}_{\ell_-}\left(\frac{P_\perp^2 \Sigma^2_{\perp}}{2}\right)\right]^2\,,
	\quad  C_0 = \frac{2^{2\ell_-} (\ell_-!)^2}{(|\ell_1|+|\ell_2|)!}\frac{\Sigma^2_{\perp}}{\pi} \,.
	\label{W0-result}
\end{equation}
For the zero impact parameter case, the absence of any external vector renders $W_0$ azimuthally symmetric. 
Moreover, for $\ell_1 + \ell_2 \neq 0$, the scattering intensity vanishes at $\bP_\perp = 0$.
This is due to the fact that, in the vicinity of this point, the transverse integral exhibits a phase vortex,
as expected from the total angular momentum conservation.

The radial dependence of $W_0$ shows different patterns for the same-sign and opposite-sign OAM choices.
If $\ell_1$ and $\ell_2$ are of the same signs, we find $\ell_- = 0$,
and since $L_0^k = 1$ for any $k$, we observe a non-oscillating dependence:
\begin{equation}
	W_0(\ell_1\ell_2 \ge 0) = \frac{\Sigma^2_{\perp}}{\pi (|\ell_1+\ell_2|)!} 
 	\cdot (P^2_\perp \Sigma^2_{\perp})^{|\ell_1+\ell_2|} \,
	e^{-P_\perp^2 \Sigma^2_{\perp}}\,.
	\label{W0-result-2}
\end{equation}
Oscillations arise only for $\ell_1$ and $\ell_2$ of opposite signs.
The origin of this feature becomes clear from an intermediate step in the computation of ${\cal I}_{0\perp}$:
it involves an $\br_\perp$ integration of the product of the coordinate space LG wave functions,
which contains $r_\perp^{|\ell_1| + |\ell_2|} e^{i(\ell_1 + \ell_2)\varphi_r}$.
If the two exponents match, $|\ell_1| + |\ell_2| = |\ell_1 + \ell_2|$, the outcome
leads to a non-oscillating function, which coincides with the transverse part of the LG wave packet \eqref{LG-general}.
If they do not match, the $r_\perp$ and $\varphi_r$ integrations result in oscillations.

All these observations are consistent with the findings of \cite{Zhao:2023cwd} and provide an analytical confirmation
of the regularities observed there.
For $\ell_2 = 0$, the process reduces to the LG-Gaussian wave packet collision,
which was treated in \cite{Li:2024mzd,Liu:2025jei}. 
$W_0$ given by the same Eq.~\eqref{W0-result-2} with $\ell_2 = 0$
coincides with the results presented there.

\subsection{Transverse momentum distribution: analytical results for $b \neq 0$}
\label{section-analytical}

\begin{figure}[!h]
	\centering
	\includegraphics[width=0.4\textwidth]{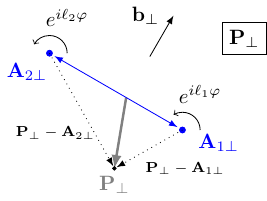}
	\caption{The auxiliary vectors $\bA_{1\perp}$ and $\bA_{2\perp}$ defined in Eq.~\eqref{vectors-A} 
		that define the positions of the phase vortices in the $\bP_\perp$-plane for the same-sign case. 
		For a generic $\bP_\perp$, the vectors $\bP_\perp - \bA_{1\perp}$ and $\bP_\perp - \bA_{2\perp}$ are also shown.}
	\label{fig-geom}
\end{figure}

Next, we move to the non-zero impact parameter case: $b \neq 0$. 
The transverse integral is more involved but still can be done analytically,
see details in Appendix~\ref{appendix-Iperp}. These expressions reveal several features with clear physics consequences. 

First, as expected, the function $W_0(\bP_\perp)$ no longer displays azimuthal symmetry.
The strongest $\varphi_P$-dependence arises for the impact parameters $b \sim \sigma_\perp$.
For the same-sign $\ell_1$ and $\ell_2$ (we choose $\ell_1, \ell_2 \ge 0$), the expression for $W_0(\bP_\perp)$ 
takes the following form:
\begin{equation}
	W_0(\ell_1,\ell_2 > 0) = C_b\, e^{-P^2\Sigma_\perp^2}\, \left[P^2+\frac{b^2}{\sigma^4_{2\perp}}+\frac{2bP}{\sigma^2_{2\perp}}\,\sin(\varphi_P-\varphi_b)\right]^{\ell_1}\,
	\left[P^2+\frac{b^2}{\sigma^4_{1\perp}}-\frac{2bP}{\sigma^2_{1\perp}}\,\sin(\varphi_P-\varphi_b)\right]^{\ell_2}\,,\label{W0-3}
\end{equation}
where $C_b$ is the normalization coefficient ensuring that $\int W_0 d^2 P_\perp = 1$. 
This remarkable form admits a simple geometrical interpretation illustrated by Fig.~\ref{fig-geom}.
For a given direction of $\bb_\perp$, let us define the momentum-space vectors $\bA_{1\perp}$ and $\bA_{2\perp}$, both orthogonal to $\bb_\perp$:
\begin{equation}
	A_1 = \frac{b}{\sigma^2_{2\perp}}\,, \quad \varphi_{A_1} = \varphi_b - \frac{\pi}{2}\quad\mbox{and}\quad
	A_2 = \frac{b}{\sigma^2_{1\perp}}\,, \quad \varphi_{A_2} = \varphi_b + \frac{\pi}{2}\,.
	\label{vectors-A}
\end{equation}
These two vectors are back-to-back but can have different lengths.
Then, $W_0$ can be written as
\begin{equation}
	W_0(\ell_1,\ell_2 > 0) = C_b \, e^{-P^2\Sigma_\perp^2} \cdot \left[(\bP_\perp - \bA_{1\perp})^{2}\right]^{\ell_1}
	\cdot \left[(\bP_\perp - \bA_{2\perp})^{2}\right]^{\ell_2}\,.
	\label{W0-geom}
\end{equation} 
This expression makes it clear that, for $\ell_1$ and $\ell_2$ of the same sign and a non-zero $\bb_\perp$, 
the distribution $W_0$ contains two zero-intensity points at $\bP_\perp = \bA_{1\perp}$ and  $\bP_\perp = \bA_{2\perp}$.
These zeros appear because these two points represent the phase vortices in the $\bP_\perp$ space, 
with the corresponding winding numbers $\ell_1$ and $\ell_2$, as confirmed by Eq.~\eqref{I0-prep-final-same}. 
Note that since the vectors $\bA_{1\perp}$ and $\bA_{2\perp}$ are back-to-back,
the only way for these two vortices to merge is precisely $b = 0$, in which case we recover Eq.~\eqref{W0-result-2}.

In the case of the opposite signs of $\ell_1$ and $\ell_2$,
we recover only one vortex at $\bP_\perp = \pm \bA_{1\perp}$ or $\pm \bA_{2\perp}$, depending on 
the signs and the magnitudes of $\ell_1$ and $\ell_2$,
with the winding number $\ell_1+\ell_2$, see Eq.~\eqref{I0-prep-final-opposite}.
In addition, a similar associated Laguerre polynomial is now present in the expression,
which produces strong intensity oscillations in the direction orthogonal to $\bb_\perp$.
For example, for $\ell_1 \le 0$ and $\ell_2 \ge 0$, the intensity has the form
\begin{eqnarray}
	 \ell_2 \ge |\ell_1|: &\quad& W_0 = C'_b\, e^{-P^2\Sigma_\perp^2} \cdot \left[(\bP_\perp - \bA_{2\perp})^{2}\right]^{|\ell_1+\ell_2|} 
	 \cdot \left|L_{|\ell_1|}^{|\ell_1+\ell_2|}(\xi)\right|^2\,,\nonumber\\[2mm]
	 \ell_2 \le |\ell_1|: &\quad& W_0 = C''_b\, e^{-P^2\Sigma_\perp^2} \cdot \left[(\bP_\perp + \bA_{1\perp})^{2}\right]^{|\ell_1+\ell_2|} 
	 \cdot \left|L_{\ell_2}^{|\ell_1+\ell_2|}(\xi)\right|^2\,.\label{W0-4}
\end{eqnarray}
The argument of the associated Laguerre polynomial is
\begin{equation}
	\xi = \frac{\Sigma_\perp^2}{2}|\bP_\perp + \bA_{1\perp}| e^{-i\varphi_{PA1}}
	\cdot |\bP_\perp - \bA_{2\perp}| e^{i\varphi_{PA2}}\,,\label{xi}
\end{equation}
where $\varphi_{PA1}$ and $\varphi_{PA2}$ are the azimuthal angles of the vectors $\bP_\perp + \bA_{1\perp}$
and $\bP_\perp - \bA_{2\perp}$, respectively.
For a generic $\bP_\perp$, the variable $\xi$ is complex. However, if $\bP_\perp$ is orthogonal to $\bb_\perp$, 
it becomes parallel to both of $\bA_{i\perp}$, which makes the argument real:
\begin{equation}
	\bP_\perp \perp \bb_\perp: \quad \xi = \frac{\Sigma_\perp^2}{2}
	\left(\pm P_\perp - \frac{b}{\sigma_{2\perp}^2}\right)
	\left(\pm P_\perp - \frac{b}{\sigma_{1\perp}^2}\right)\,.\label{xi-2}
\end{equation}
As $P_\perp$ changes, the associated Laguerre polynomial displays oscillations.

In addition, if $\sigma_{1\perp} = \sigma_{2\perp}$, we obtain $\bA_{1\perp} = -\bA_{2\perp}$,
and the argument $\xi$ becomes real for all $\bP_\perp$. 
The zeros of the polynomial are reaches not at isolated points but on the circles of specific radii. 

We arrive at the conclusion that a non-zero impact parameter $b$ leads to two remarkable effects for scattering of two LG vortex wave packets.
For the opposite-sign $\ell_1$ and $\ell_2$, we expect to observe a new, non-radial interference pattern, 
distinct from the radial oscillations discussed above.
For the same-sign $\ell_1$ and $\ell_2$, we predict that the initial phase vortex 
splits in the $\bP_\perp$-space into two well-separated vortices with the winding numbers $\ell_1$ and $\ell_2$.
The momentum-space separation between the vortices is governed by the impact parameter $\bb_\perp$.
We stress that these two vortices arises in the momentum distribution of the total final two-particle system, 
not of individual final particles.

\subsection{Transverse momentum distribution: numerical examples}

\begin{figure}[h]
	\centering
	\includegraphics[width=0.32\textwidth]{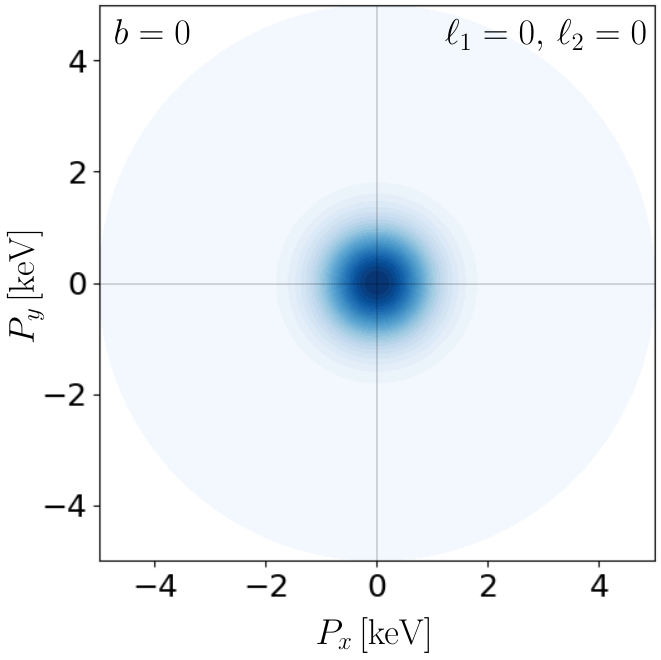}\quad
	\includegraphics[width=0.32\textwidth]{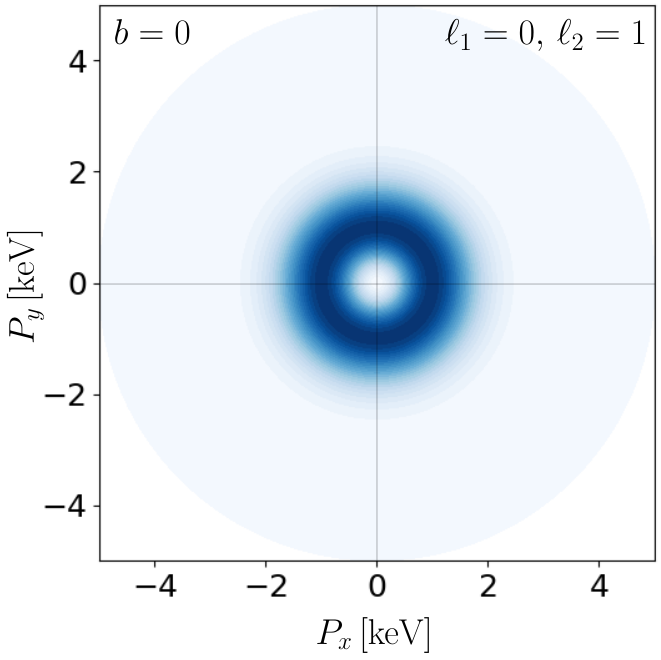}\quad
	\includegraphics[width=0.32\textwidth]{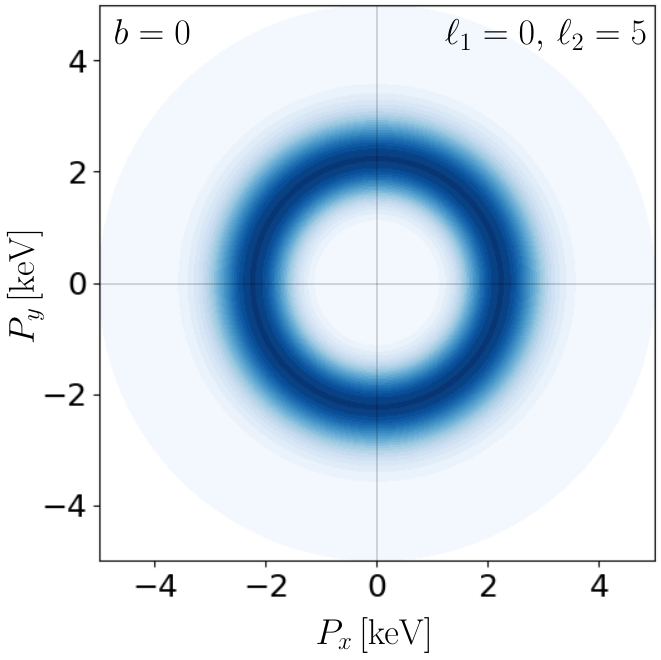}\\
	\includegraphics[width=0.32\textwidth]{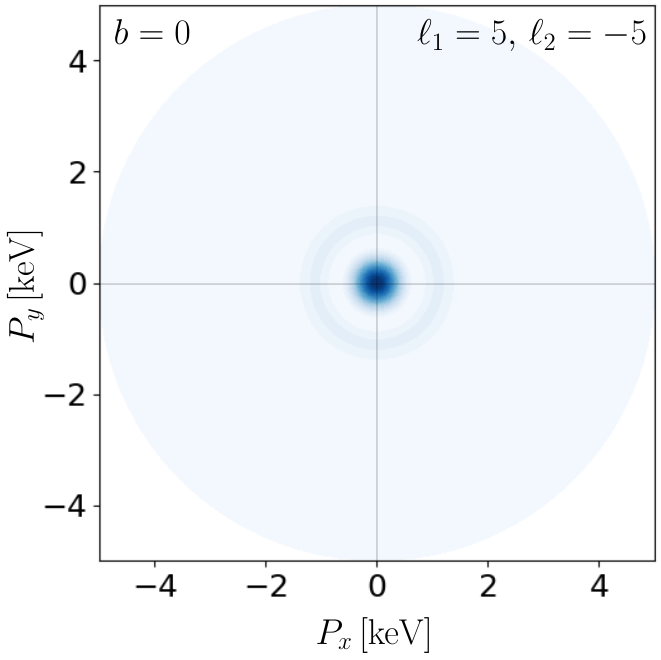}\quad
	\includegraphics[width=0.32\textwidth]{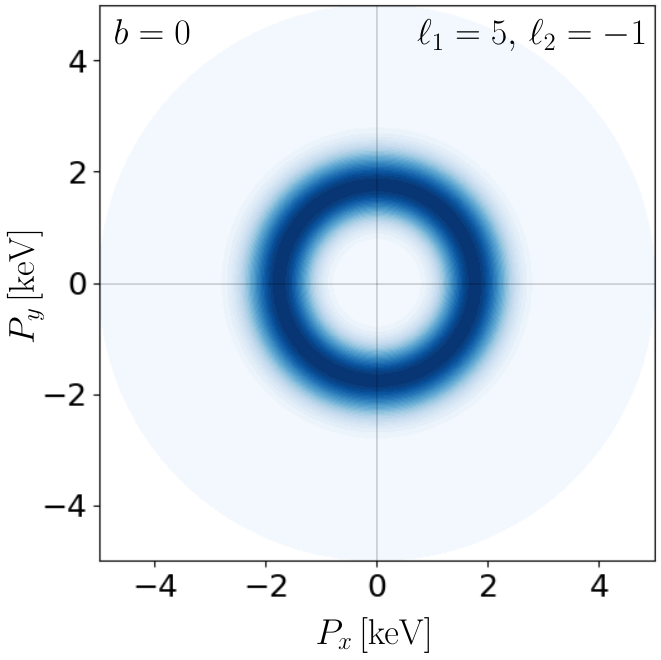}\quad
	\includegraphics[width=0.32\textwidth]{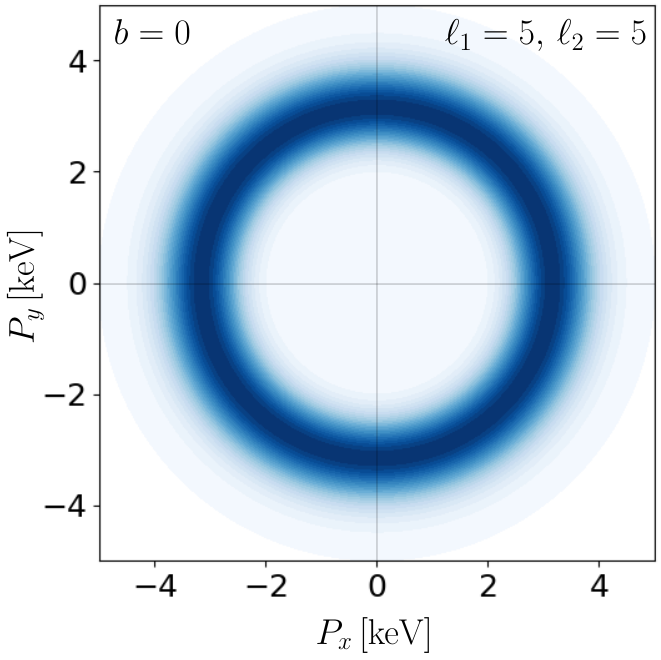}
	\caption{The function $W_0(\bP_\perp)$ shown by the shade intensity for $\Sigma_\perp = 1~\keV^{-1}$ at zero impact parameter, $b=0$. 
		Top row: $\ell_1 = 0$ and $\ell_2 = 0, 1, 5$.
		Bottom row: $\ell_1 = 5$, and $\ell_2 = -5, -1, 5$.}
	\label{fig-W0-zero-b}
\end{figure}

Let us now illustrate the above insights with specific examples.
We will show a sequence of plots for $W_0(\bP_\perp)$ drawn from the analytical expressions 
for several illustrative choices of the LG parameters. We begin with the coaxial case $b = 0$
and show results for specific values of $\ell_1, \ell_2$ such as 0, $\pm 1$, and $\pm 5$.
Note that for $b = 0$, $W_0$ depends not on $\sigma_{1\perp}$ and $\sigma_{2\perp}$ separately 
but on their combination $\Sigma_\perp$ defined in Eq.~\eqref{cal-I0perp}.
For concreteness, we choose the tightly focused wave packets 
with $\sigma_{1\perp} = \sigma_{2\perp} = \sqrt{2}\,\keV^{-1} \approx 0.3\,\nm$ corresponding to 
$\Sigma_\perp = 0.2\,\nm = 1\,\keV^{-1}$, so that
the typical extent of the $\bP_\perp$ distribution is in the keV range. Note that such a tight sub-nm-range focusing 
of vortex electrons has already been demonstrated in experiment \cite{Verbeeck:2011-atomic}.
For a less tight focusing, the scale of the $\bP$-distribution will shrink proportionally.

In Fig.~\ref{fig-W0-zero-b}, top row, we choose $\ell_1 = 0$ and show how $W_0$ depends on the value of $\ell_2$.
In this and all subsequent figures, the value of $W_0$ is encoded in the shade intensity; 
its numerical value is unimportant since $W_0$ always satisfies the overall normalization $\int W_0 d^2P_\perp = 1$. 
As expected, these density plots are azimuthally symmetric 
and exhibit a single ring of the radius $P_\perp^2 \Sigma^2_{\perp} = |\ell_2|$, in agreement with Eq.~\eqref{W0-result-2}.
In Fig.~\ref{fig-W0-zero-b}, bottom row, we show similar distributions for $\ell_1 = 5$ and $\ell_2 = -5, -1, 5$.
The plots follow the general trend mentioned above. For the same-sign case, we observe a single ring, while  
for opposite signs of $\ell_1$ and $\ell_2$, we begin to discern radial oscillations.
They are not very pronounced, though, as they take place in the $P_\perp$ range beyond the main Gaussian peak.
For $\ell_1 + \ell_2 = 0$, the function $W_0$ is non-zero at the origin, $\bP_\perp = 0$,
but the width of the central Gaussian peak decreases as we increase the value of $|\ell_1|$. 
These plots make it clear that observing any $\bP_\perp$ structure at $b=0$
will be much easier for the same-sign arrangement, $\ell_1 = \ell_2$, than for the opposite signs
such as $\ell_1 = -\ell_2$.
All these findings are consistent with the results of \cite{Zhao:2023cwd}.

\begin{figure}[!h]
	\centering
	\includegraphics[width=0.32\textwidth]{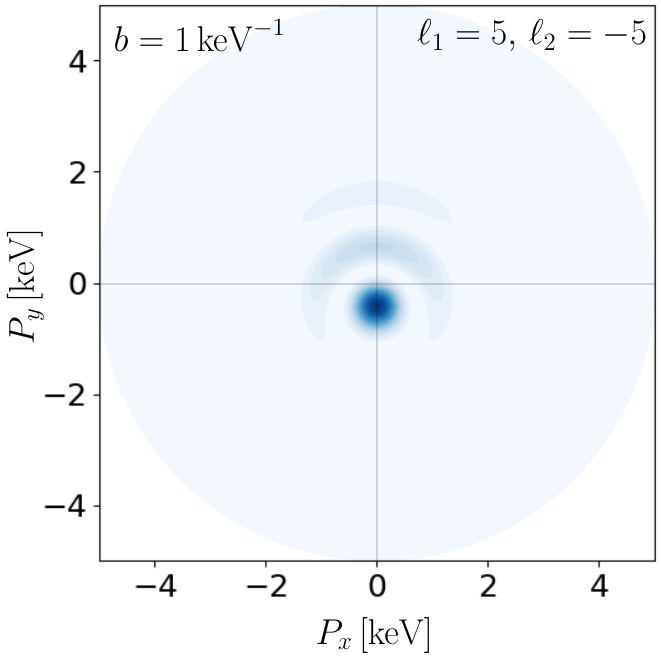}\quad
	\includegraphics[width=0.32\textwidth]{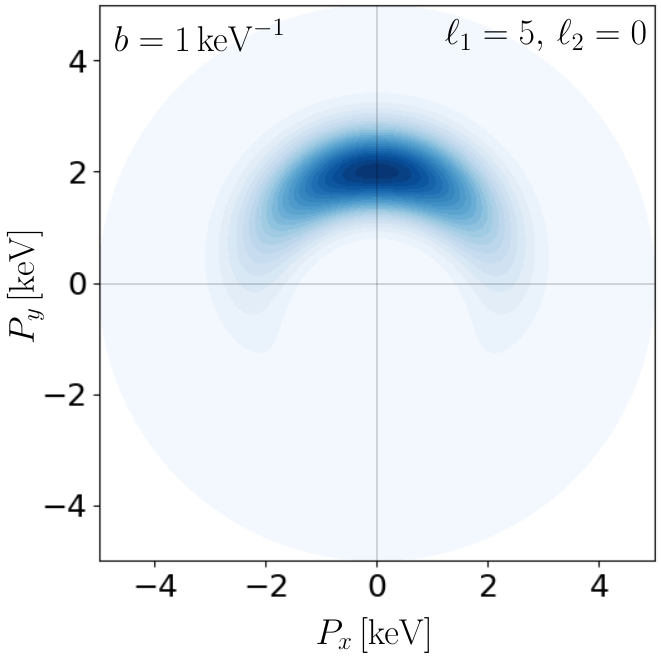}\quad
	\includegraphics[width=0.32\textwidth]{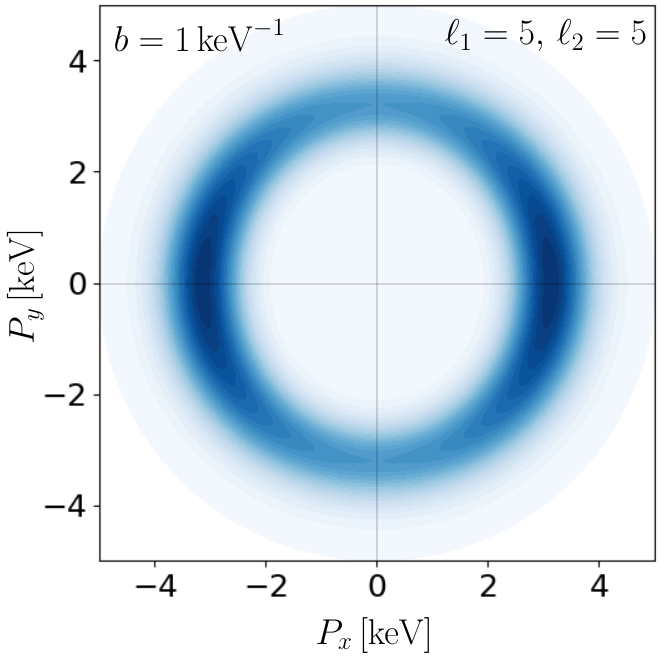}\\
	\includegraphics[width=0.32\textwidth]{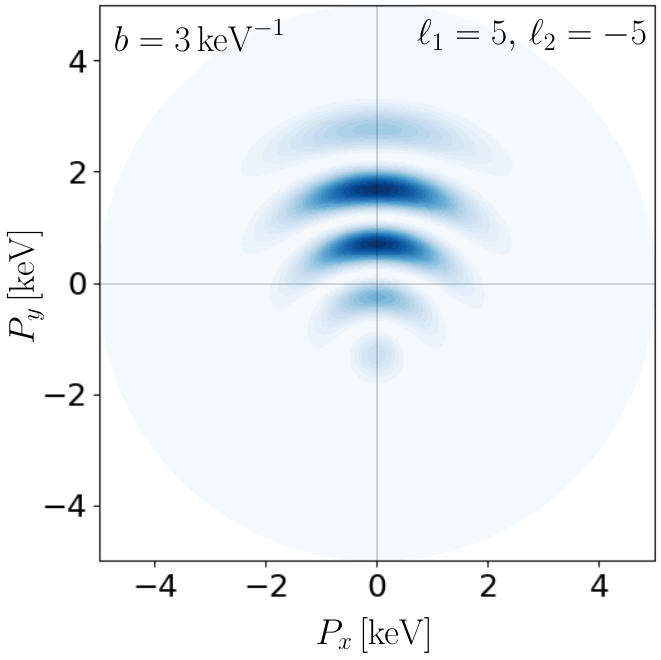}\quad
	\includegraphics[width=0.32\textwidth]{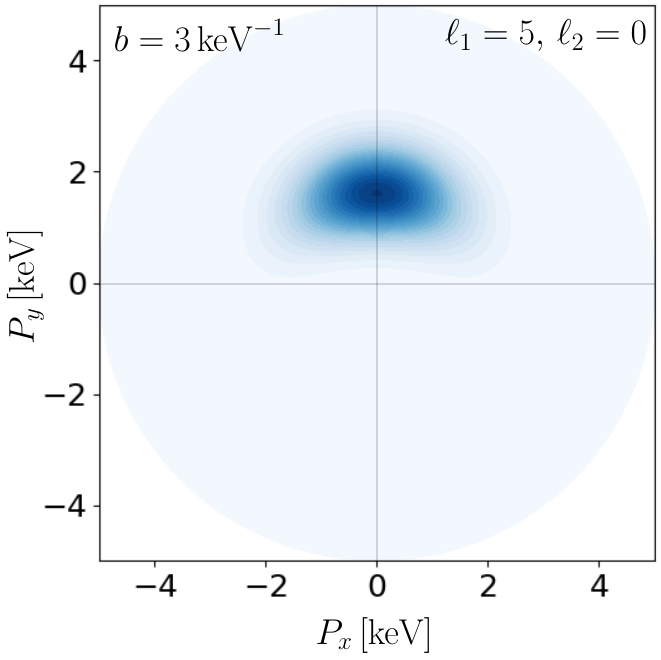}\quad
	\includegraphics[width=0.32\textwidth]{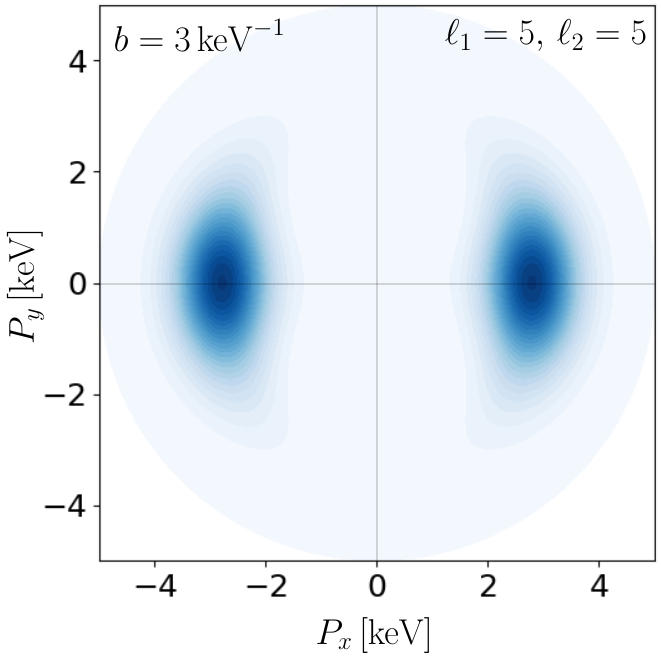}\\
	\caption{The function $W_0(\bP_\perp)$ for non-zero $b=1\,\keV^{-1}$ (top row) and $3\,\keV^{-1}$ (bottom row) 
		and for $\ell_1 = 5$ and $\ell_1 = -5$ (left), 0 (middle), and 5 (right column).
		In all cases, $\sigma_{1\perp} = \sigma_{2\perp} = 1.41\,\keV^{-1}$.}
	\label{fig-W0-nonzero-b}
\end{figure}

Next, we show the effect of a non-zero impact parameter $\bb_\perp$, which has never been systematically studied for LG-LG scattering.
For the non-vortex situation, $\ell_1 = \ell_2 = 0$, the presence of non-zero impact parameter between two Gaussian wave packets
does not change the $\bP_\perp$ distribution, and we do not show this case in the plots. 

Fig.~\ref{fig-W0-nonzero-b} presents a grand picture of the effects induced by an offset in the $x$ direction: $\bb_\perp = (b, 0)$.
In all the plots, the LG parameters are $\sigma_{1\perp} = \sigma_{2\perp} = 1.41~\keV^{-1}$.
The two rows correspond to the impact parameter $b = 1\,\keV^{-1} = 0.2\,\nm$ and $3\,\keV^{-1} = 0.6\,\nm$,
while the sequence of plots in each row refers to different $(\ell_1,\ell_2)$.
The plots in the left column correspond to the opposite-sign $(\ell_1,\ell_2) = (5,-5)$ and exhibit 
a curious interference pattern in the direction orthogonal to $\bb_\perp$, which
was anticipated in our discussion of the analytical results.
As we increase $b$, the envelop shifts from negative to positive $P_y$.
The plots in the middle column correspond to the LG-Gaussian collision with $(\ell_1,\ell_2) = (5,0)$.
We observe here the hints of the superkick effect discussed in \cite{Ivanov:2022sco,Li:2024mzd,Liu:2025jei}.
Finally, the right column corresponds to $(\ell_1,\ell_2) = (5,5)$ and exhibits the splitting 
of the single central vortex with $\ell_1+\ell_2$ into two separate vortices along the $P_y$ direction.
For $b=3~\keV^{-1}$, the two vortices are located at $P_y = 1.5~\keV$.

All these effects are in a qualitative agreement with our discussion of the analytic results.
In the next section, we will put them under scrutiny to gain better understanding of their 
dependence on the LG wave packet properties and on the impact parameter $b$. 

We close this section by mentioning that these features for the LG-LG overlap integrals
significantly differ from the corresponding Bessel-Bessel collision expressions 
\cite{Ivanov:2011kk,Ivanov:2012na,Zhao:2023cwd}. 
For collision of two Bessel states with the transverse momentum parameters 
$\varkappa_1$ and $\varkappa_2$, the relevant integral at $b = 0$ is 
\begin{eqnarray}
	{\cal J} &=& 
	\int d^2 k_{1\perp}\,d^2 k_{2\perp}\, \delta^{(2)}(\bk_{1\perp} + \bk_{2\perp} - \bP_\perp)\, 
	\delta(k_{1\perp}-\varkappa_1) \delta(k_{2\perp}-\varkappa_2) \, e^{i(\ell_1\varphi_1 + \ell_2\varphi_2)}\nonumber\\
	&=&e^{i(\ell_1+\ell_2)\varphi_P} \frac{\varkappa_1 \varkappa_2}{\Delta}\cdot \cos(\ell_1\delta_1 - \ell_2 \delta_2)\,.
	\label{Bessel-1}
\end{eqnarray}
Here, $\Delta$ is the area of the triangle with sides $P_\perp$, $\varkappa_1$, and $\varkappa_2$,
while $\delta_i$ are the inner angles between $P_\perp$ and $\varkappa_i$.
As $P_\perp$ changes from its smallest value $|\varkappa_1 - \varkappa_2|$ to its largest value $\varkappa_1 + \varkappa_2$,
the angles $\delta_i$ vary significantly, which drives the radial oscillations.
These oscillations are present for any signs of $\ell_1$ and $\ell_2$, although their number increases
for the opposite-sign values. These oscillations are very prominent near the edges of the $P_\perp$ ring.
Smearing the exact Bessel states with a Gaussian $\varkappa$ dependence helps 
suppress the highest frequency oscillations and avoid the sharp edges; 
nevertheless, the oscillations persist for any signs of $\ell_1$ and $\ell_2$.
In our case of the principal mode LG-LG scattering, the radial oscillations are absent in the same-sign case
and are weak for the opposite-sign cases unless we rely on a sizable impact parameter $b$.

\section{Consequences of a non-zero impact parameter}\label{section-insights}

In this section, we will present a more in-depth study of several physical effects mentioned above.
What unites all of them is the key role played by the non-zero impact parameter.
Thus, the key message of this work is that in vortex-vortex scattering,
a non-zero impact parameter should be regarded not as a nuisance parameter but 
as an additional powerful tool that helps reveal the full physics potential of vortex state scattering.

\subsection{Transverse momentum imbalance: a paradox and its resolution}

The non-trivial, unbalanced $\bP_\perp$-distributions of $W_0$ shown in Fig.~\ref{fig-W0-nonzero-b} may seem paradoxical.
Indeed, since each initial LG wave packet has $\lr{\bk_{i\perp}} = 0$,
the average total initial momentum is zero: $\lr{\bk_{1\perp} + \bk_{2\perp}} = 0$.
Due to momentum conservation, we expect to see a net zero total final transverse momentum in the final state as well.
This is indeed the case for collision of two Gaussian wave packets, $\ell_1 = \ell_2 = 0$, even at a non-zero $b$.
However, Fig.~\ref{fig-W0-nonzero-b} takes it clear that for vortex-Gaussian and vortex-vortex collisions with $\ell_1 \neq \ell_2$,
the average total final momentum is non-zero: $\lr{\bP_\perp} \neq 0$.
This apparent momentum non-conservation may lead the reader to suspect a flaw in our calculations.

\begin{figure}[!h]
	\centering
	\includegraphics[width=0.35\textwidth]{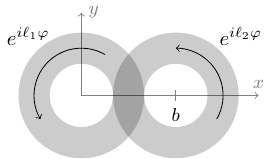}
	\caption{Schematic coordinate-space overlap of two LG wave packets offset by impact parameter $b$.
	The overlap region leads to most intense scattering and leads to a non-zero total final state momentum $P_y$.}
	\label{fig-overlap}
\end{figure}

In reality, this counter-intuitive feature is one of the hallmark peculiarities of vortex state scattering.
It was explicitly exposed and resolved in \cite{Liu:2022nfq} for vortex-Gaussian scattering; 
the same logic applies to the vortex-vortex case. For the reader's convenience, we repeat here the resolution of this paradox
for the LG-LG scattering.

The short answer is that the total momentum is, of course, conserved, and the average final state transverse momentum 
is zero---provided we take into account the {\em entire} final state. 
In Fig.~\ref{fig-overlap}, we show the two LG wave packets in coordinate space at the moment of their collision in the focal plane.
The wide rings of radius $\rho_i = \sqrt{|\ell_i|}\sigma_{i\perp}$ schematically represent their donut-shape transverse intensity profiles.
For simplicity, Fig.~\ref{fig-overlap} is drawn for $\rho_1 = \rho_2$.
Inside each ring, the azimuthal phase gradient leads to the local tangential momentum 
of the order of $2\pi |\ell_i|/2\pi \rho_i = \sqrt{|\ell_i|}/\sigma_{i\perp}$.
This can also be seen through the Wigner function for the paraxial LG state \cite{Karlovets:2020odl}.
The impact parameter $b$ is chosen in Fig.~\ref{fig-overlap} along the $x$ direction, and its magnitude is approximately equal
to the sum of the two radii. 
The region of the largest overlap, which is shown by a darker shade of gray,
contains the total local momentum along the $y$ direction of the order of
\begin{equation}
	P_y \approx \mbox{sign}(\ell_1) \frac{\sqrt{|\ell_1|}}{\sigma_{1\perp}} 
	- \mbox{sign}(\ell_2)\frac{\sqrt{|\ell_2|}}{\sigma_{2\perp}}\,.\label{biased-local-momentum} 
\end{equation}
A typical collision event that leads to a sizable deflection of {\em each} of the two incoming particles originates predominantly 
from this overlap region. Therefore, the total transverse momentum of the {\em substantially} scattered final state is biased towards $P_y$. 
This is what we expect to find as we detect the scattered particles and measure they total transverse momentum.

However, these deflected pairs of particles do not represent the {\em entire} final state. 
In addition to the wave function component describing the particles scattered away from their respective vortex cones,
there remains the part of the total wave function that escaped this hard scattering. Being outside of the overlap region,
it carries the compensating total momentum, which averages to $-P_y$. However, each particle in this ``non-scattered'' component
of the final state stays approximately within the corresponding vortex ring, and as a result, it avoid detection.

The end result is that the entire final state, which includes the detected outcome of a hard scattering
and the non-detected components, has net transverse momentum zero. However, since we only detect a part of this final state,
we seem to obtain a biased non-zero $P_y$. Thus, a non-zero $\lr{\bP_\perp}$ appears because it is a {\em conditional} average
over all detected events, not over the full final state.

It must be stressed that this peculiar transverse momentum imbalance is characteristic of vortex state scattering.
It does not emerge for collision of two Gaussian states, let alone two plane waves.
Experimental detection of such a momentum imbalance will represent
a novel quantum effect induced by the vortex states and a non-zero impact parameter.
It can also be used as a diagnostic tool for high-energy vortex state detection.

\subsection{High-contrast interference}

\begin{figure}[!h]
	\centering
	\includegraphics[width=0.23\textwidth]{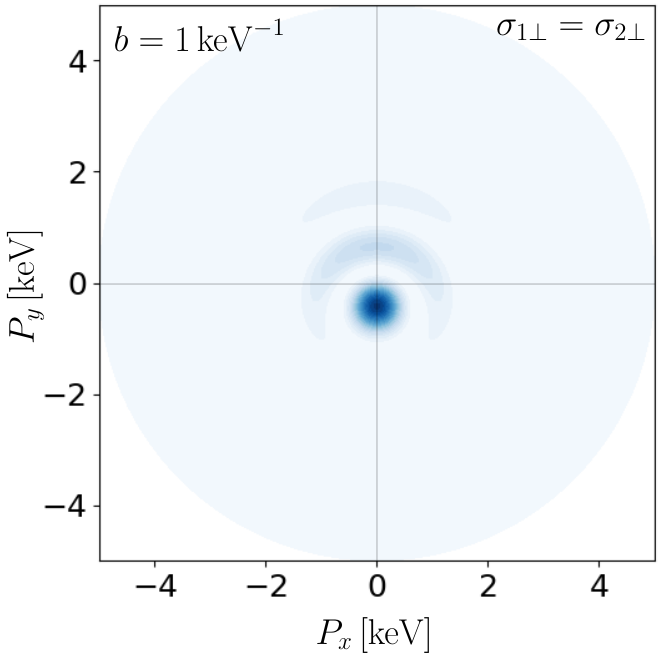}\quad
	\includegraphics[width=0.23\textwidth]{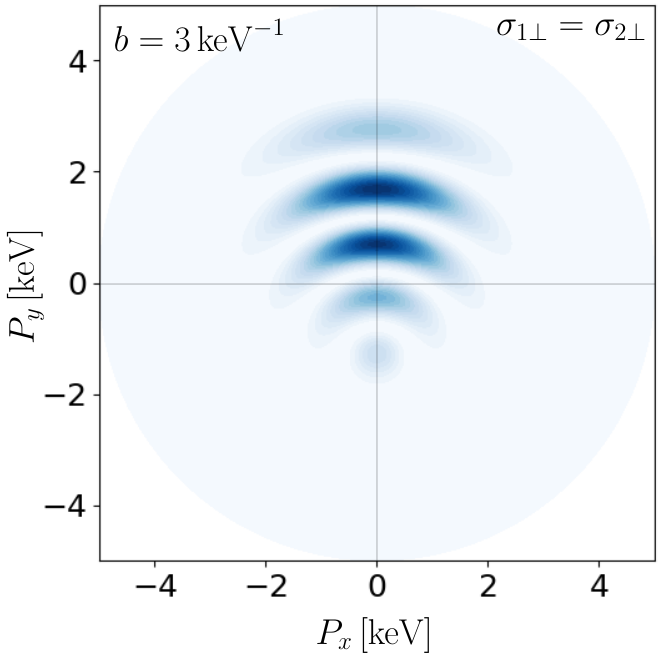}\quad
	\includegraphics[width=0.23\textwidth]{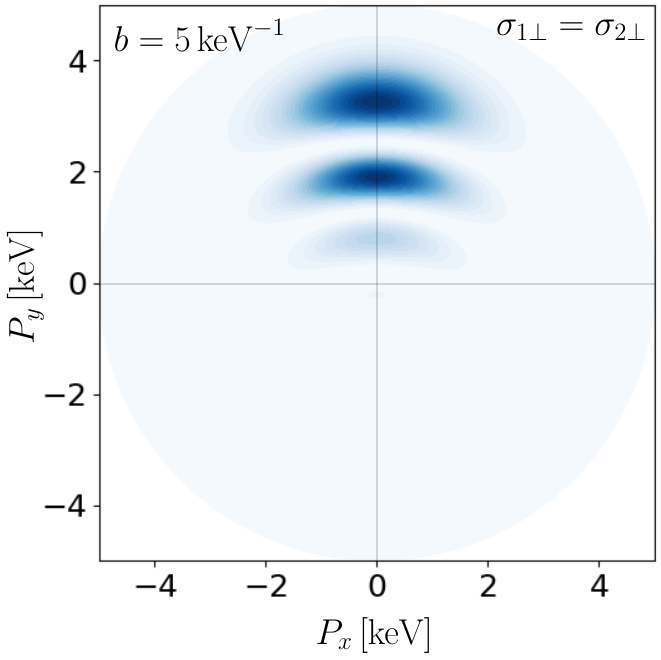}\quad
	\includegraphics[width=0.23\textwidth]{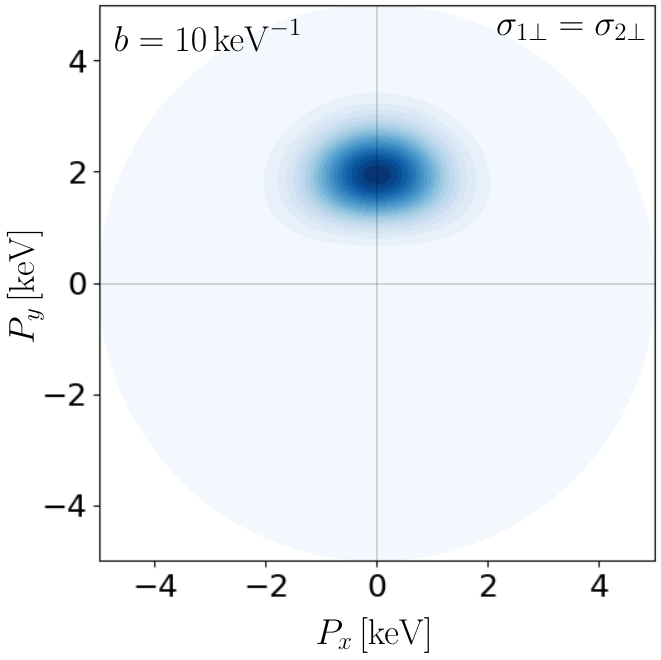}\\
	\includegraphics[width=0.23\textwidth]{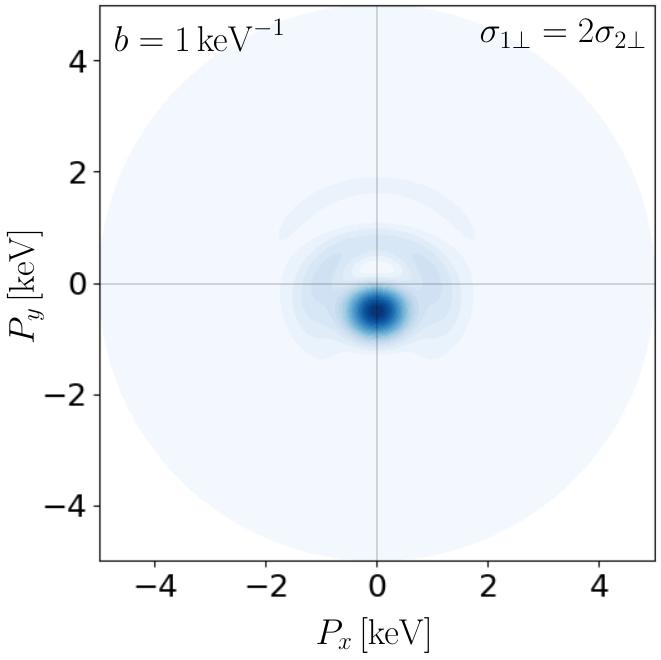}\quad
	\includegraphics[width=0.23\textwidth]{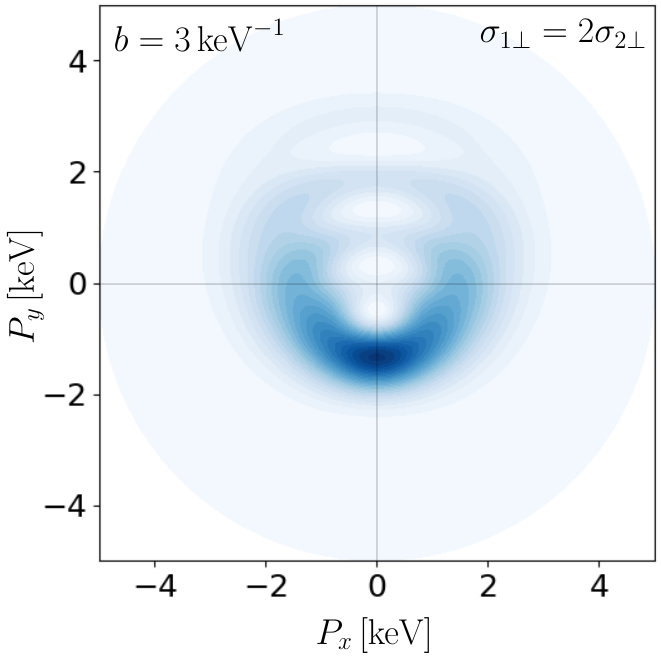}\quad
	\includegraphics[width=0.23\textwidth]{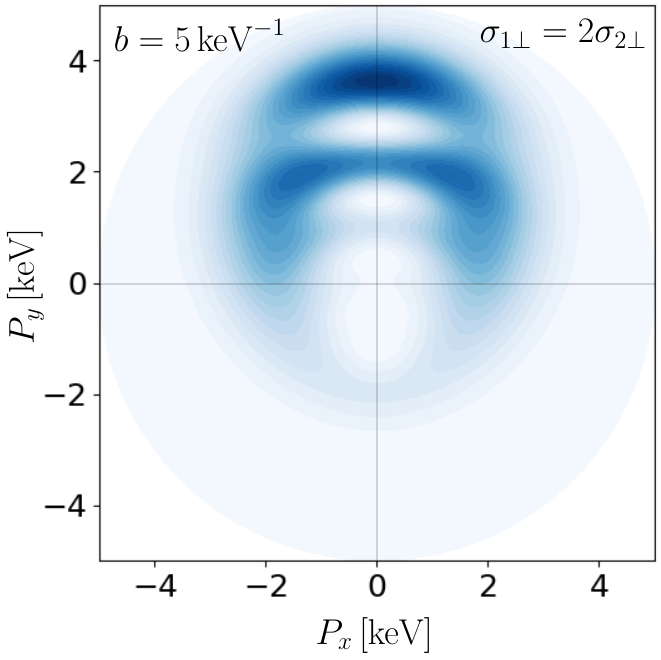}\quad
	\includegraphics[width=0.23\textwidth]{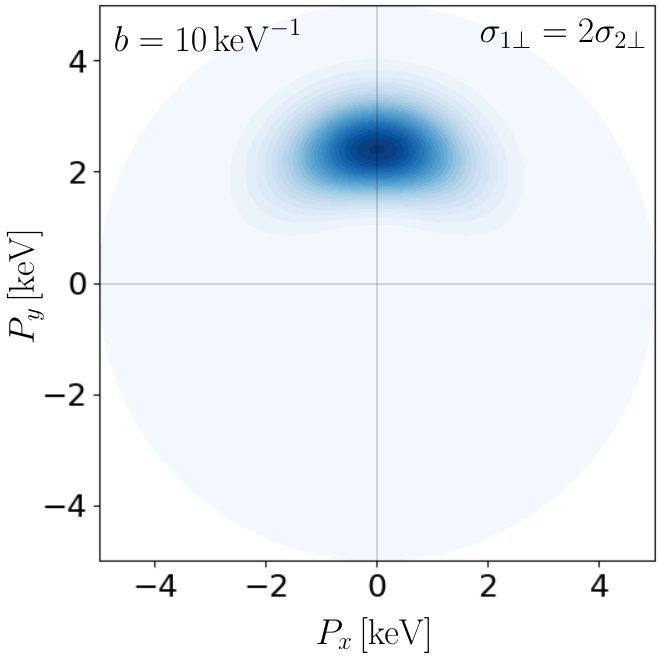}
	\caption{Top: the function $W_0(\bP_\perp)$ for the opposite-sign OAM $(\ell_1,\ell_2) = (5,-5)$ as a function of the impact parameter 
		for $b = 1, 3, 5$, and $10\,\keV^{-1}$.
		The top row corresponds to $\sigma_{1\perp} = \sigma_{2\perp} = 1.41~\keV^{-1}$, 
		the bottom row corresponds to $\sigma_{1\perp} = 2\sigma_{2\perp} = 2~\keV^{-1}$.}
	\label{fig-W0-high-contrast}
\end{figure}

As discussed in the previous section, for $\ell_1$ and $\ell_2$ of opposite signs, the intensity $W_0$ exhibits oscillations, 
which originate from interference between different initial plane wave configurations producing the same final state.
For the coaxial case $b=0$, the oscillations are weak, as they were located in the $P_\perp$ region where $W_0$
is already suppressed by the Gaussian factor $\exp(-\Sigma_\perp^2 P_\perp^2)$.
A non-zero $b$ effectively shifts these oscillations in the central $P_\perp$ region, making them very pronounced.

In Fig.~\ref{fig-W0-high-contrast}, we show $W_0(\bP_\perp)$ for several values of the impact parameter $b = 1, 3, 5$, and $10\,\keV^{-1}$,
and for two choices of the LG wave packet sizes: $\sigma_{1\perp} = \sigma_{2\perp} = 1.41~\keV^{-1}$ (top row)
and $\sigma_{1\perp} = 2\sigma_{2\perp} = 2~\keV^{-1}$ (bottom row).
As $b$ increases, the oscillations develop strong azimuthal angle-dependence, aligning in the direction orthogonal to $\bb_\perp$.
The strongest contrast is seen in the second plot of the top row of Fig.~\ref{fig-W0-high-contrast}, 
which we call the ``wifi plot'' due to its similarity with the iconic wifi symbol.
For even higher values of $b$, the argument $\xi$ in Eq.~\eqref{xi-2} stays large
in the central $\bP_\perp$ region, and the oscillations die out.
Note also that choosing sufficient different $\sigma_{1\perp}$ and $\sigma_{2\perp}$, such as 
$\sigma_{1\perp} = 2\sigma_{2\perp}$ in the bottom row of Fig.~\ref{fig-W0-high-contrast},
leads to additional distortions of the oscillations. 
Thus, if one aims to detect a clean oscillation pattern as in the ``wifi'' plot, one should strive 
for $\sigma_{1\perp} = \sigma_{2\perp}$.

\subsection{The superkick effect}

In 2013, Barnett and Berry theoretically discovered the surprising phenomenon of superkick
in the interaction of localized probes with vortex light \cite{barnett2013superkick}, 
see also recent developments in \cite{AlDrees2026}.
Imagine a trapped atom interacting with the vortex light and located at a non-zero impact parameter $b$
from the phase vortex axis.
If the vortex light beam is described by a LG beam with some $\ell$ and $\sigma_\perp$,
then the typical photon transverse momentum is $\varkappa \approx \sqrt{|\ell|}/\sigma_\perp$.
Now, if $b\ll \sigma_\perp$, the phase gradient probed by the atom is $|\ell|/b \gg \varkappa$. 
Thus, when the atom absorbs a vortex photon, it can receive a transverse recoil 
much bigger than the transverse momentum of any photon in the light beam, hence the superkick.

This phenomenon was later considered for hadronic photoproduction 
and deuteron photo-disintegration \cite{Afanasev:2020nur,Afanasev:2021fda},
where it was predicted to significantly alter the energy threshold and the cross section behavior.
These predictions were critically reexamined in \cite{Ivanov:2022sco,Liu:2022nfq} for generic two particle scattering
in LG-Gaussian kinematics. This effect was proposed as a diagnostic tool for high-energy vortex electrons \cite{Li:2024mzd,Liu:2025jei}
and as a promising route to the first ever detection of laboratory manifestation of non-linear QED effects in light-light scattering \cite{Bu:2025mer}. 

	Although the superkick effect has not yet been observed experimentally, the situation may change soon.
In \cite{Gaudout:2025nnd}, the recent experimental progress was reported in probing the local momentum 
distribution and the phase front of a shaped laser beam with the aid of an atom interferometer 
employing a compact Bose-Einstein condensate as a momentum recoil sensor.
So far, only the longitudinal momentum recoil was measured; the authors of \cite{Gaudout:2025nnd}
detected an extra recoil at the ppb scale. Since the laser beam was non-vortex,
this extra recoil was driven by the intensity gradient, with phase fluctuations playing a minor role.
It will be interesting to see if this interferometric scheme
can be adapted to detection of the transverse momentum recoil in vortex lasers, where the azimuthal phase gradient 
can impart a much stronger transverse superkick discussed here.

The superkick effect in light-atom interaction may seem paradoxical in the semiclassical picture, 
in which the target is considered as a classical localized probe, but it finds its resolution
in the fully quantum picture that treats the target as a wave packet \cite{barnett2013superkick}
that contains PW components with sufficiently large momenta.
In free-space LG-Gaussian scattering, it was also resolved within the QFT approach, \cite{Ivanov:2022sco,Liu:2022nfq}.
Our present calculations corresponds to the original superkick effect when $\ell_1 \neq 0$, $\ell_2 = 0$ 
and $\sigma_{2\perp} \ll \sigma_{1\perp}$.
We can now verify if the effect survives for non-zero $\ell_i$.

\begin{figure}[!h]
	\centering
	\includegraphics[width=0.3\textwidth]{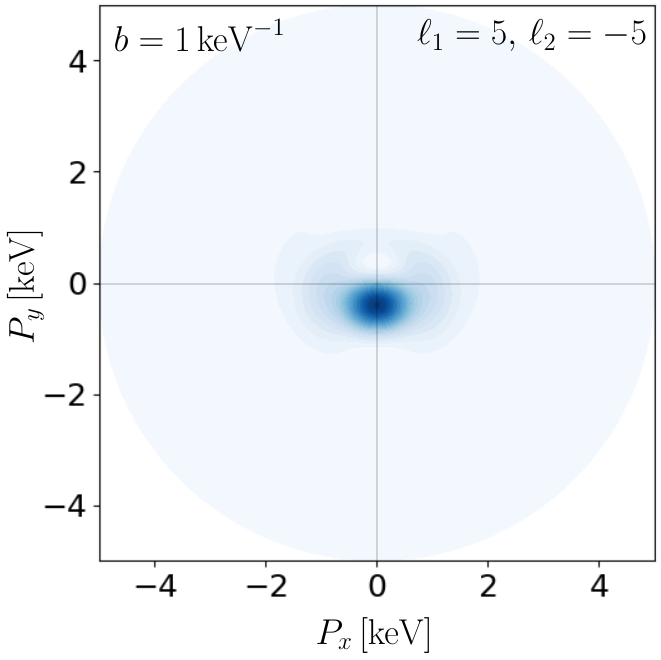}\quad
	\includegraphics[width=0.3\textwidth]{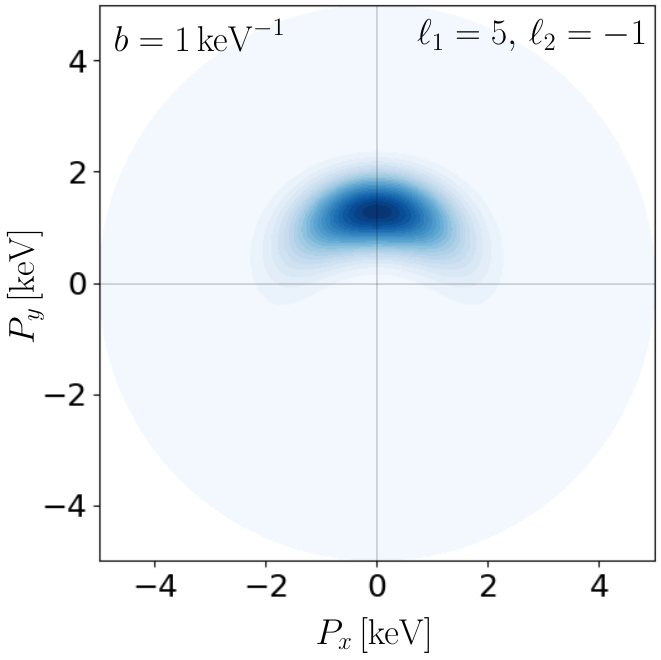}\quad
	\includegraphics[width=0.3\textwidth]{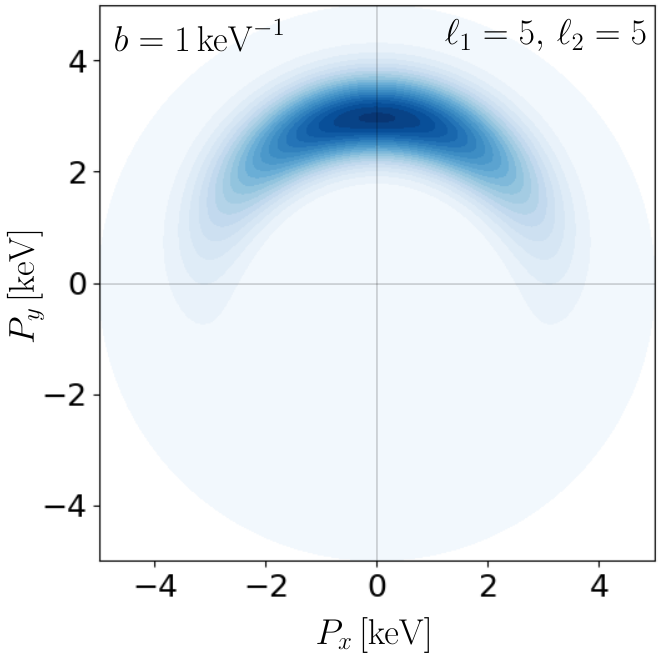}\\
	\caption{The function $W_0(\bP_\perp)$ for $(\ell_1,\ell_2) = (5,-5), (5,-1), (5,5)$ for $b = 1~\keV^{-1}$
		and $\sigma_{1\perp} = 10 \sigma_{2\perp} = 10~\keV^{-1}$.}
	\label{fig-superkick}
\end{figure}

In Fig~\ref{fig-superkick}, we show the intensity distribution $W_0(\bP_\perp)$ 
for $(\ell_1,\ell_2) = (5,-5)$, $(5,-1)$, and $(5,5)$. The transverse sizes of the wave packets
are chosen $\sigma_{1\perp} = 10 \sigma_{2\perp} = 10~\keV^{-1}$, 
and the impact parameter is $b = \sigma_{2\perp} = 1~\keV^{-1}$. 
We clearly see that the total momentum transfer significantly overshoots the typical
transverse momentum of the wider wave packet $\sqrt{\ell_1}/\sigma_{1\perp} = 0.23~\keV^{-1}$,
the hallmark feature of the superkick effect. Of course, its emergence
in two wave packet collisions of strongly unequal $\sigma_{i\perp}$ is no surprise.

\subsection{Vortex splitting}

\begin{figure}[!h]
	\centering
	\includegraphics[width=0.3\textwidth]{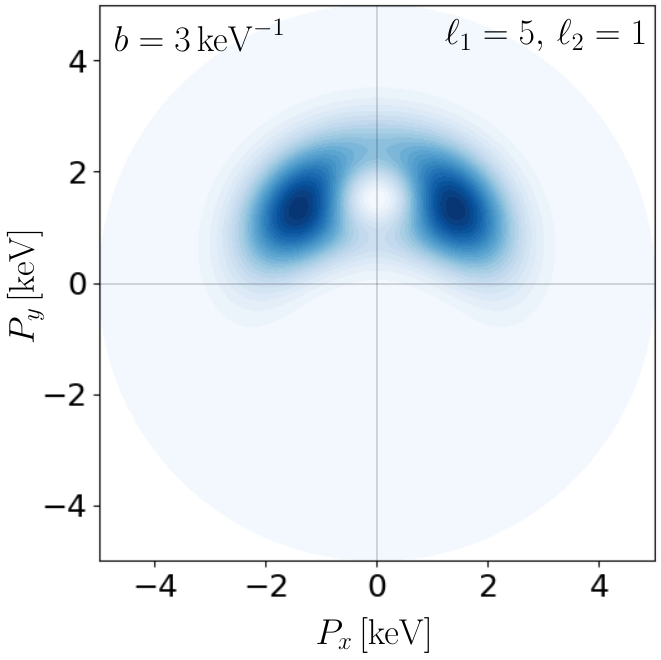}\quad
	\includegraphics[width=0.3\textwidth]{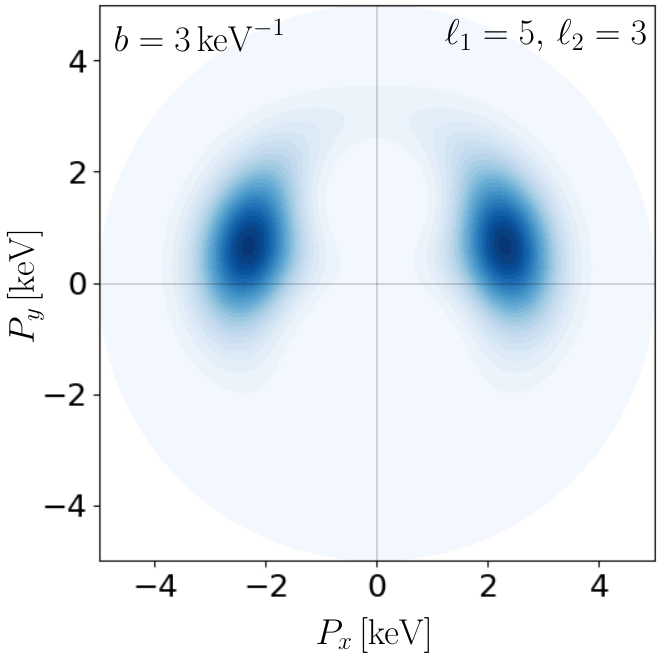}\quad
	\includegraphics[width=0.3\textwidth]{b3-5-5}\\
	\caption{Splitting of the initial vortex into two vortices 
		with the parameters $\sigma_{1\perp} = \sigma_{2\perp} = 1.41~\keV^{-1}$, $b=3~\keV^{-1}$,
		and $(\ell_1,\ell_2) = (5,1)$, $(5,3)$, and $(5,5)$. }
	\label{fig-splitting}
\end{figure}

As discussed in Section~\ref{section-analytical}, for the case of same-sign $\ell_1$ and $\ell_2$,
a non-zero $b$ leads to the remarkable effect of a single vortex splitting into
two vortices with the winding numbers $\ell_1$ and $\ell_2$.
The analytical expressions Eq.~\eqref{W0-geom} make it clear that the visibility of splitting depends
on all the parameters of the collision. In Fig.~\ref{fig-splitting}, we illustrate this effect for 
$\sigma_{1\perp} = \sigma_{2\perp} = 1.41~\keV^{-1}$, $b=3~\keV^{-1}$,
and $(\ell_1,\ell_2) = (5,1)$, $(5,3)$, and $(5,5)$. 
With these parameters, the two vortices are located at $P_y = \pm b/\sigma_{i\perp}^2 = \pm1.5~\keV$,
which are visible in the plots. Note that in the first two plots the ``hole'' at $P_y = - 1.5~\keV$ is stronger
than at $P_y = 1.5~\keV$ because it corresponds to the $\ell_1 > \ell_2$.

This vortex-splitting effect offers an interesting opportunity for vortex-state experiments.
We can set up LG-LG scattering with fixed $(\ell_1,\ell_2)$ and a fixed $b$.
After sizable statistics of scattering events has been accumulated, we can select two sub-samples
centered in the $\bP_\perp$ plane around the first and the second momentum space vortices.
This will allow us to explore production of two-particle final states with $\ell_{tot} \approx \ell_1$
and with $\ell_{tot} \approx \ell_2$ in the {\em same collision settings}, and therefore with the same systematic uncertainties,
without the need to change the initial $\ell_i$.
It remains to be explored what insights into particle structure and interactions such a setting can bring.
What we stress here is that a non-zero controllable impact parameter $b$ allows for such an experimental protocol.

\section{Discussion and outlook}\label{section-discussion}


Collisions of vortex states of high-energy photons, electrons, protons, neutrons, and other particles 
represent a novel experimental tool for nuclear and particle physics.
The intrinsic orbital angular momentum carried by vortex states, their topologically protected plane-wave superposition, 
and the peculiar spin-OAM entangled polarization opportunities lead to effects 
that cannot be mimicked in plane wave or Gaussian wave packet scattering.
Numerous predictions of such effects still await their experimental verification \cite{Ivanov:2022jzh}.
However, the calculations leading to these predictions were done under very diverse assumptions.
In particular, the implicit requirements on the initial state preparation and final state detection
often go far beyond the capabilities of possible near-future experiments.
All this makes it difficult to systematically compare theoretical predictions among themselves 
and to judge on the prospects of their experimental verification.

In this paper, we initiate a systematic re-analysis of vortex state scattering 
under uniform and, hopefully, realistic assumptions.
We work in the general relativistic kinematics, not limiting ourselves to the non-relativistic or ultra-relativistic approximations.
We do adopt certain assumptions for the parameters of the colliding vortex states,
which we believe offer a reasonable approximation to the future high-energy vortex states of electrons, photons, or protons.
\begin{enumerate}
	\item 
	We consider free-space collisions of two paraxial LG wave packets in their principal modes.
	The key parameters are the OAM projections $\ell_i$ and transverse localization parameters $\sigma_{i\perp}$
	of the two wave packets, as well as the controllable impact parameter $\bb_\perp$ between the two vortex axes.
	\item 
	We describe the final state particles as plane waves with momenta $\bk_1'$ and $\bk_2'$, to be detected with conventional pixelized detectors with a suitable resolution.
	\item 
	Calculations are done under the impulse approximation, in which the spreading of the wave packets during the collision event is neglected. 
\end{enumerate}
We followed the standard procedure for wave packet scattering 
developed in \cite{Kotkin:1992bj} and refined recently in \cite{Karlovets:2020odl}.
The usage of the well-defined LG wave packets eliminates all the artefacts originating from the unnormalized Bessel beams
\cite{Ivanov:2011kk} and avoids the somewhat arbitrary Gaussian-smearing procedure adopted in \cite{Ivanov:2011bv} and in later works.

For $2\to 2$ plane-wave scattering in the center of mass frame, $\bk_1' + \bk_2'= 0$,
and the cross section can be expressed as $d\sigma(\bk_{1}')$.
For wave packet scattering, the $2\to 2$ differential cross section becomes a function of both final momenta,
$d\sigma(\bk_1',\bk_2')$, or as  $d\sigma(\bk_1',\bP)$, where $\bP = \bk_1' + \bk_2'$.
Repeating the point made in \cite{Ivanov:2012na,Ivanov:2016oue}, we argued that the main novelty of vortex-vortex scattering 
comes not from the $\bk'_1$-dependence but from the $\bP$-dependence of $d\sigma(\bk'_1,\bP)$, 
in particular from its $\bP_\perp$-distribution.
Integrating $d\sigma(\bk'_1,\bP)$ over all total momenta will result in a cross section 
nearly identical to the plane wave cross section $d\sigma(\bk_{1\perp}')$.

Hence the first take-home message of this study: the benefits of vortex-vortex scattering come not from numerical correction 
to the plane wave scattering but from the new final phase space dimensions that open up for exploration.
There is no counterpart for such dimensions in plane-wave scattering.

Next, we stress that $\bP_\perp$-dependent features of $d\sigma(\bk'_1,\bP)$ can originate both from
the initial wave packet arrangements and from the properties of the scattering amplitude itself.
The former source is ``instrumental'' and universal;
it depends on how we prepare the colliding vortex states. 
The latter source is what we usually aim to study,
as it may reveal non-trivial interaction dynamics, especially in hadronic processes.
For a safe interpretation of any peculiarity that could arise in future studies of $d\sigma(\bk'_1,\bP)$,
it is imperative to disentangle these two sources.
This conclusion is the second key message of our work.

This is why we focused in this work on the former source alone. We defined the function $W_0(\bP_\perp)$
normalized to $\int W_0(\bP_\perp) d^2 P_\perp = 1$ that reveals the universal kinematic features
of the total transverse momentum distributions arising in two LG wave packet collisions.
We found the exact analytic expressions for $W_0$ for general $\ell_i$, $\sigma_{i\perp}$, and the impact parameter $\bb_\perp$.
These expressions reveal several remarkable physics effects:
a seemingly paradoxical transverse momentum imbalance, 
high-contrast interference fringes that emerge at non-zero impact parameter $b$,
and a controllable splitting, in the $\bP_\perp$ space, 
of a single initial $\ell_1+\ell_2$ vortex into two vortices with the winding numbers $\ell_1$ and $\ell_2$.
All these effects are new, specific to vortex states, and have never been detected in particle scattering experiments.
Their verification in elastic electron-electron scattering will represent an important reality check 
for vortex scattering experiments.

When exploring these effects, we repeatedly saw the instrumental role of the impact parameter $b$.
Usually, a non-zero impact parameter is considered an undesired effect in vortex state scattering.
In our study, its role changes completely; we now argue that a controllable impact parameter is a powerful tool 
for revealing vortex-induced effects rather than a nuisance factor. 
This conclusion represents the third message of our work.

One may wonder how the function $W_0$ at a fixed $\bP_\perp$ depends on $\bb_\perp$.
The answer comes from the remark in section~\ref{section-general-properties}: 
the integral ${\cal I}_{0\perp}$ has the same structure in momentum and coordinate space,
with the roles of $\bP_\perp$ and $\bb_\perp$ swapped. Therefore, impact parameter 
dependence of $W_0$ at fixed $\bP_\perp$ mirrors the $\bP_\perp$ distribution at fixed $\bb_\perp$.
The combinations $\bP_\perp \pm \bA_{i\perp}$ in our analytic results confirm this expectations.

The effects that we described are observable already with the present-day technology,
with vortex electrons of momentum $p = 600~\keV$, 
which corresponds to the kinetic energy of $300~\keV$ typical for electron microscopes.
Suppose two such electrons are brought in collision, scatter at small-to-moderate angles, 
and are detected in coincidence by pixelized electron detectors with pixel size of half a millimeter
installed one meter downstream. The angular resolution of the detected electrons is then $0.5\times 10^{-3}$.
This translates into the total momentum resolution of the order of $0.3~\keV$,
which is more than enough for all the effects we discussed.
Even if focusing of the initial vortex electrons is ten times less tight, in the nm range,
one can still resolve all the effects with the electron detector pixels sizes of $100~\mu m$,
which is feasible today. The visibility of the effects also depends on how stable the impact parameter $b$ is.
One can model the effect of $\bb_\perp$ jitter by smearing the $W_0(\bP_\perp)$ plots 
with a certain Gaussian function; the resulting distributions will of course lose their contrast.
However, if $b$ jitter is less than $\sigma_{i\perp}$, which is certainly the case in the present-day electron microscopes,
one can expect the main effects stay.
Thus, we argue all these novel features can be experimentally verified with the existing technology
if two electron microscope sources of vortex electrons are mounted in a collider-like fashion.

The results of this work open the road to the future systematic study of specific non-resonant processes
such as $ee \to ee$, the Compton scattering, and electron-proton elastic or inelastic scattering.
It will also allow us to treat scattering of initial state electrons or photons in exotic spin-OAM entangled polarization states,
which are unavailable for plane wave scattering.
Delegated to a future work is also an extension of the present technique to resonant processes, 
in which the plane-wave scattering amplitude varies sharply within the wave packet momentum integration domain;
that formalism will allow us to update the Bessel-state analysis of \cite{Ivanov:2019pdt,Ivanov:2019vxe} with LG scattering.

\section*{Acknowledgments}

We are indebted to Nikolay Volchansky, Dmitry Karlovets, Dmitry Naumov, Liangliang Ji, and Anton Murtazin for valuable comments and helpful discussions.
This work was supported by the Fundamental Research Funds for the Central Universities, Sun Yat-sen University, China.


\appendix

\section{Impulse approximation}\label{appendix-impulse}

Localized wave packets are not monochromatic and evolve in time.
The exact expressions for $\psi(\br, t)$ for LG wave packets, both within and beyond the paraxial approximation, 
were already presented in \cite{Karlovets:2018iww,Karlovets:2020odl, Liu:2022nfq}.
We quote these results, in the paraxial approximation, in the notation of \cite{Liu:2022nfq}:
\begin{equation}
	\psi_1(\br, t) = \pi^{-3/4} \frac{i^{\ell_1}}{\sqrt{|\ell_1|!}}\frac{\sqrt{\sigma_{1\perp}^2\sigma_{1z}}}{\tilde\sigma_{1\perp}^2\tilde\sigma_{1z}}
	\left(\frac{\sigma_{1\perp} r_\perp}{\tilde\sigma_{1\perp}^2}\right)^{|\ell_1|} e^{i\ell_1\varphi_r}
	\exp\left[-\frac{r_\perp^2}{2\tilde\sigma_{1\perp}^2}-\frac{(z-v_1 t)^2}{2\tilde\sigma_{1z}^2}\right]\, e^{-i\varepsilon_1 t + i p_{1z}z}\,.
	\label{LG-general-r}
\end{equation}
Here, we used the shorthand notation for the complex-valued time-dependent combinations
\begin{equation}
	\tilde\sigma_{1\perp}^2 = \sigma_{1\perp}^2 + i \frac{t}{\varepsilon_1}\,, \quad 
	\tilde\sigma_{1z}^2 = \sigma_{1z}^2 + i \frac{t}{\gamma_1^2 \varepsilon_1}\,.\label{tilde-sigma-1}
\end{equation}
The probability density then takes the following form
\begin{equation}
	|\psi_1(\br, t)|^2 = \frac{1}{\pi^{3/2}|\ell_1|!} \, \frac{1}{\sigma_{1\perp}^2(t) \sigma_{1z}^2(t)}
	\left(\frac{r_\perp^2}{\sigma_{1\perp}^2(t)}\right)^{|\ell_1|}\exp\left[-\frac{r_\perp^2}{\sigma_{1\perp}^2(t)}
	-\frac{(z-v_1 t)^2}{\sigma_{1z}^2(t)}\right]\,, 
\end{equation} 
where the effective time-dependent localization lengths are
\begin{equation}
	\sigma_{1\perp}^2(t) \equiv \sigma_{1\perp}^2 \left(1 + \frac{t^2}{\sigma_{1\perp}^4\varepsilon_1^2}\right)\,,\quad
	\sigma_{1z}^2(t) \equiv \sigma_{1z}^2 \left(1 + \frac{t^2}{\sigma_{1z}^4\gamma_1^4 \varepsilon_1^2}\right)\,.\label{sigma-perp-z-t}
\end{equation}
We recovered the well-known spreading of the wave packet as it propagates,
with the typical spreading time being $\sigma_{1\perp}^2\varepsilon_1$ for the transverse dynamics 
and $\sigma_{1z}^2\gamma_1^2\varepsilon_1$ for the longitudinal one.
For a wave packet with $\sigma_{1z} = \sigma_{1\perp}/\gamma_1$, the two brackets in Eqs.~\eqref{sigma-perp-z-t} are equal, 
and the spreading wave packet preserves its shape. 

The time evolution of the counter propagating LG state follows the similar pattern but includes the impact parameter and 
the offset of the focal point:
\begin{equation}
	\psi_2(\br, t) = \pi^{-3/4} \frac{i^{\ell_2}}{\sqrt{|\ell_2|!}}\frac{\sqrt{\sigma_{2\perp}^2\sigma_{2z}}}{\tilde\sigma_{2\perp}^2\tilde\sigma_{2z}}
	\left(\frac{\sigma_{2\perp} r_\perp}{\tilde\sigma_{2\perp}^2}\right)^{|\ell_2|} e^{i\ell_2\varphi_r}
	\exp\left[-\frac{(\br_\perp-\bb_\perp)^2}{2\tilde\sigma_{2\perp}^2}-\frac{(z-v_2 t)^2}{2\tilde\sigma_{2z}^2}\right]\, e^{-i\varepsilon_2 t + i p_{2z}z}\,.
\end{equation}
where
\begin{equation}
	\tilde\sigma_{2\perp}^2 = \sigma_{2\perp}^2 + i \frac{t}{\varepsilon_2}\,, \quad 
	\tilde\sigma_{2z}^2 = \sigma_{2z}^2 + i \frac{t}{\gamma_2^2 \varepsilon_2}\,.\label{tilde-sigma-2}
\end{equation}

These explicit expressions can now be inserted into the luminosity integral \eqref{lumi}.
However, instead of computing the integrals exactly, we remark that the main contribution comes from the time interval
of the order of $t_c = \sqrt{\sigma_{1z}^2+ \sigma_{2z}^2}/|v_1-v_2|$,
which we call the duration of the collision event.
The crucial step now is to assume that the longitudinal and transverse localization scales 
$\sigma_{i\perp}$ and $\sigma_{iz}$ do not change significantly during the collision event.
This condition is satisfied automatically for the longitudinal scale due to
$\sigma_{iz}p_{iz} \gg 1$, 
so that the main constraint comes in the form of an upper limit on $\sigma_{1z}^2 + \sigma_{2z}^2$:
\begin{equation}
	\sqrt{\sigma_{1z}^2 + \sigma_{2z}^2} < \sigma_{i\perp}^2 \varepsilon_i |v_1-v_2| \,.\label{impulse-conditions}
\end{equation}
We call this assumption the {\em impulse approximation}.
Put simply, we avoid dealing with wave packets that are too long for a given transverse localization scale.

Once we adopt the impulse approximation, we can replace $\sigma_i(t) \to \sigma_i$ taken at the collision momentum.
This leads to simplifications in the wave packet scattering amplitude ${\cal I}$ defined in Eq.~\eqref{cal-I}.
Following \cite{Liu:2022nfq}, we first convert the energy delta-function $\delta(E_1+E_2-E_f)$ into time integration:
\begin{equation}
	{\cal I} = \frac{1}{(2\pi)^7} \int_{-\infty}^{+\infty} dt\, e^{iE_ft}
	\int\frac{d^3k_1 d^3k_2}{4E_1E_2}\, e^{-i(E_1 + E_2)t}  \phi_1(\bk_1)\,\phi_2(\bk_2)
	\delta^{(3)}(\bk_{1}+\bk_{2} - \bP)\,\cdot {\cal M}\,.\label{cal-I-2}
\end{equation}
As we work in the paraxial approximation, $|\bk_{i\perp}|$, $|\bP_\perp|$, 
as well as $\delta k_{iz} = k_{iz}-p_{iz}$ are much smaller than $|p_{iz}|$.
We do not limit ourselves to the center of motion frame,
which means that $P_z = k_{1z}' + k_{2z}'$ can be large but $\Delta P_z = P_z - (p_{1z} + p_{2z})$
is small, $|\Delta P_z| \ll |p_{iz}|$.
Then, under the impulse approximation, we get
\begin{equation}
	E_1 + E_2 \approx \varepsilon_1 + \varepsilon_2 + v_1 \delta k_{1z} + v_2 (\Delta P_z - \delta k_{1z})\,.
\end{equation}
We also consider a generic kinematics for final state momenta so that the plane-wave amplitude ${\cal M}$
is a smooth function and its dependence on $k_{1z}$ can be neglected within the small $\delta k_{1z}$ range.
Then performing the integrals as described in the appendix of \cite{Liu:2022nfq},
we arrive to the following expression for the vortex scattering amplitude:
\begin{equation}
	{\cal I} =  \frac{1}{(2\pi)^4\, \sqrt{\varepsilon_1\varepsilon_2|v_1-v_2|}} 
	\frac{\sigma_{1\perp}\sigma_{2\perp}}{\sqrt{|\ell_1|!\,|\ell_2|!}}  
	\cdot {\cal I}_L \cdot {\cal I}_\perp\,.\label{cal-I-3}
\end{equation}
Here, 
\begin{eqnarray}
	{\cal I}_L = \frac{\sqrt{\sigma_{1z}\sigma_{2z}}}{\sqrt{\pi\,|v_1-v_2|}} 
	\, \exp{\left[ -\frac{1}{2}(\Delta P_z)^2\frac{\sigma_{1z}^2\sigma_{2z}^2}{\sigma_{1z}^2+\sigma_{2z}^2} 
		-\frac{\sigma_{1z}^2 + \sigma_{2z}^2}{2(v_1-v_2)^2} \left(  \delta E - \Delta P_z \frac{v_1 \sigma_{2z}^2 + v_2 \sigma_{1z}^2}{\sigma_{1z}^2 + \sigma_{2z}^2}  \right)^2  \right] }\label{I-L-goodbye}
\end{eqnarray}
is a Gaussian function of the final state kinematic variables $\Delta P_z = P_z - (p_{1z} + p_{2z})$ 
and $\delta E \equiv E_f- \varepsilon_1 - \varepsilon_2$. 
The prefactor in ${\cal I}_L$ was chosen in such a way that $\int |{\cal I}_L|^2 dP_z dE_f = 1$.
The transverse part is
\begin{eqnarray}
	{\cal I}_\perp &=& 
	\int d^2 k_{1\perp}\,d^2 k_{2\perp}\, \delta^{(2)}(\bk_{1\perp} + \bk_{2\perp} - \bP_\perp)\, 
	(\sigma_{1\perp}k_{1\perp})^{|\ell_1|} \, (\sigma_{2\perp}k_{2\perp})^{|\ell_2|}  \, e^{i(\ell_1\varphi_1 + \ell_2\varphi_2)}\nonumber\\
	&&\quad \times \ \exp\left[ -\frac{k_{1\perp}^2 \sigma_{1\perp}^2}{2} -\frac{k_{2\perp}^2 \sigma_{2\perp}^2}{2} 
	- i\bb_\perp \bk_{2\perp}\right]	\cdot {\cal M}\,.\label{I-perp2}
\end{eqnarray}
It is this quantity ${\cal I}_\perp$ that is of particular interest to us, as it contains the all-important interference patterns
in the total transverse momentum $\bP_\perp$.

\section{The transverse integral ${\cal I}_{0\perp}$}\label{appendix-Iperp}

\subsection{Zero impact parameter: $b = 0$}

We start with case of the zero impact parameter. The transverse integral given in Eq.~\eqref{I0-redef} becomes
\begin{equation}
	{\cal I}_{0\perp} =
	\int d^2 k_{1\perp}\,d^2 k_{2\perp}\, \delta^{(2)}(\bk_{1\perp} + \bk_{2\perp} - \bP_\perp)\, 
	(\sigma_{1\perp}k_{1\perp})^{|\ell_1|} \, (\sigma_{2\perp}k_{2\perp})^{|\ell_2|}  \, e^{i(\ell_1\varphi_1 + \ell_2\varphi_2)}
\cdot \exp\left[ -\frac{k_{1\perp}^2 \sigma_{1\perp}^2}{2} -\frac{k_{2\perp}^2 \sigma_{2\perp}^2}{2} \right]	\,.\label{I0-perp-ap}
\end{equation}
The two-dimensional delta function can be recast in the form the a coordinate space integral
\begin{equation}
(2\pi)^2\delta^{(2)}(\bk_{1\perp} + \bk_{2\perp} - \bP_\perp) = \int d^2 r_\perp\, \exp[i(\bk_{1\perp} + \bk_{2\perp} - \bP_\perp) \cdot \br_\perp]\,.
\end{equation}
It allows us to factorize the momentum space integrations:
\begin{eqnarray}
	{\cal I}_{0\perp} &=& 
	\frac{1}{(2\pi)^2} \cdot \int d^2 r_\perp\cdot e^{-i\bP_\perp \cdot \br_\perp} \int d^2 k_{1\perp}\, 
	(\sigma_{1\perp}k_{1\perp})^{|\ell_1|} \cdot \exp\left[-\frac{k_{1\perp}^2 \sigma_{1\perp}^2}{2}+i\bk_{1\perp} \cdot \br_\perp + i\ell_1\varphi_1\right]\nonumber\\
	&&\quad \int d^2 k_{2\perp}\,(\sigma_{2\perp}k_{2\perp})^{|\ell_2|} \cdot \exp\left[-\frac{k_{2\perp}^2 \sigma_{2\perp}^2}{2}+i\bk_{2\perp} \cdot \br_\perp + i\ell_2\varphi_2\right]\,.\label{I0-perp-wodelta-ap}
\end{eqnarray}
Using $\int d\varphi_1 \, e^{i\ell_1 \varphi_1}\,  e^{i\bk_{1\perp} \cdot \br_\perp}=2\pi i^{|\ell_1|} e^{i\ell_1\varphi_r}J_{|\ell_{1}|}(rk_{1\perp})$, 
where $r \equiv |\br_\perp|$, and Eq.~(6.631.4) from \cite{gradshteyn2014table}, we obtain  
\begin{equation}
I_1 = \int d^2 k_{1\perp}\, 
(\sigma_{1\perp}k_{1\perp})^{|\ell_1|} \cdot \exp\left[-\frac{k_{1\perp}^2 \sigma_{1\perp}^2}{2}+i\bk_{1\perp} \cdot \br_\perp + i\ell_1\varphi_1\right]
=\frac{2\pi i^{|\ell_1|}}{(\sigma_{1\perp})^{|\ell_1|+2}}\
\cdot r^{|\ell_1|}\, e^{-\frac{r^2}{2\sigma_{1\perp}^2}+ i\ell_1\varphi_r}\, ,
\end{equation}
The evaluation reduces then to a single remaining integration: 
\begin{equation}
\begin{split}
\mathcal{I}_{0\perp} &=\frac{i^{|\ell_1|+|\ell_2|}}{\sigma_{1\perp}^{|\ell_1|+2} \sigma_{2\perp}^{|\ell_2|+2}}\int dr\cdot r^{|\ell_1|+|\ell_2|+1}\ e^{-\frac{r^2}{2\Sigma^2_{\perp}}} \int d\varphi_r \cdot e^{i(\ell_1+\ell_2) \varphi_r}\  e^{-i\bP_\perp \cdot \br_\perp} \\
&=2\pi\,\frac{(-1)^{\ell_-}}{\sigma_{1\perp}^{|\ell_1|+2} \sigma_{2\perp}^{|\ell_2|+2}}\ e^{i(\ell_1+\ell_2)\varphi_P} \int dr\cdot r^{|\ell_1|+|\ell_2|+1}\ e^{-\frac{r^2}{2\Sigma^2_{\perp}}}J_{|\ell_1+\ell_2|}(P_{\perp}r)\,,
\end{split}
\end{equation}
where $\Sigma^2_{\perp}$ and $\ell_-$ were defined in Eq.~\eqref{cal-I0perp} and Eq.~\eqref{ell-minus} and we repeat the definitions here:
$$
\Sigma^2_{\perp} \equiv \frac{\sigma^2_{1\perp}\sigma^2_{2\perp}}{\sigma^2_{1\perp} + \sigma^2_{2\perp}}\,,
\quad 
\ell_- \equiv \frac{|\ell_1| + |\ell_2| - |\ell_1 + \ell_2|}{2} = 
\left\{
\begin{array}{l}
	0 \quad \mbox{for $\ell_1\cdot \ell_2 \ge 0$,} \\[2mm]
	\min(|\ell_1|, |\ell_2|) \quad \mbox{for $\ell_1 \cdot \ell_2 < 0$.}
\end{array}	
\right.
$$
By using Eqs.~(6.631.10) from \cite{gradshteyn2014table}, 
we get our final result:
\begin{equation}
	\mathcal{I}_{0\perp} = \frac{2\pi}{\sigma_{1\perp}^2 + \sigma_{2\perp}^2} (-2)^{\ell_-}\, \ell_-! \cdot \frac{(\Sigma^2_{\perp})^{|\ell_1+\ell_2| + \ell_-}}{\sigma^{|\ell_1|}_{1\perp}\sigma^{|\ell_2|}_{2\perp}}\cdot P_{\perp}^{|\ell_1+\ell_2|}\, e^{i(\ell_1+\ell_2)\varphi_P} \exp\left(-\frac{P_\perp^2 \Sigma^2_{\perp}}{2}\right) \cdot L^{|\ell_1+\ell_2|}_{\ell_-}\left(\frac{P_\perp^2 \Sigma^2_{\perp}}{2}\right)\,.
	\label{I0perp-zero-b}
\end{equation}
A quick check of dimensions: the integral $\mathcal{I}_{0\perp}$ is of dimension $\sigma^{-2}$, 
which is taken care by the first factor, while the $\sigma$-dimensions of the remaining factors 
sum up to $2|\ell_1+\ell_2| + 2\ell_- - |\ell_1+\ell_2| - |\ell_1| - |\ell_2| = 0$.

\subsection{Non-zero impact parameter: $b \not = 0$, same-sign $\ell_1, \ell_2$}

For non-zero $\bb_\perp$, switching to the coordinate space integration can also be done in the same way, 
with the result shown in Eq.~\eqref{I0-redef-r}.
However, it does not bring us any simplification. Thus, we start again from the definition Eq.~\eqref{I0-redef} and follow a different strategy.
First, we perform the $\bk_{2\perp}$ integration:
\begin{eqnarray}
	{\cal I}_{0\perp} &=& \sigma_{1\perp}^{|\ell_1|} \sigma_{2\perp}^{|\ell_2|} e^{-i\bb_\perp \cdot \bP_\perp}
	\int d^2 k_{1\perp}\, k_{1\perp}^{|\ell_1|} e^{i\ell_1\varphi_1}\; 
	|\bP_\perp - \bk_{1\perp}|^{|\ell_2|}  \, e^{i\ell_2\varphi_{Pk1}}\nonumber\\
	&&\qquad\qquad\qquad \times \ \exp\left[ -\frac{k_{1\perp}^2 \sigma_{1\perp}^2}{2} -\frac{(\bP_\perp - \bk_{1\perp})^2 \sigma_{2\perp}^2}{2} 
	+ i\bb_\perp \cdot \bk_{1\perp}\right]	\,,\label{I0-redef-2}
\end{eqnarray}
where $\varphi_{Pk1}$ is the azimuthal angle of the vector $\bP_\perp - \bk_{1\perp}$.
Next, we express all the complex quantities $k e^{i\varphi} = k_x + ik_y$ via scalar products
with suitable ``helicity'' vectors: 
\begin{equation}
	k_{1\perp} e^{i\varphi_1} \equiv \bk_{1\perp}\cdot \bbe_+\,, \quad
	k_{1\perp} e^{-i\varphi_1} \equiv \bk_{1\perp}\cdot \bbe_-\,, \quad
	\bbe_+ = \doublet{1}{i}\,, \quad \bbe_- = \doublet{1}{-i}\,.
\end{equation}
These vectors satisfy the properties $\bbe_+ \cdot \bbe_+ = \bbe_- \cdot \bbe_- = 0$
and $\bbe_+ \cdot \bbe_- = 2$.

Already at this stage, the expressions begin to depend on the signs of $\ell_1$ and $\ell_2$.
Let us now choose $\ell_1 > 0$ and $\ell_2 > 0$ and proceed with the calculation.

Next, we now represent the factors such as $(k_{1\perp} e^{i\varphi_1})^{\ell_1}$
as arising from repeated differentiation with respect to real-valued parameters $t_1, t_2$:
\begin{equation}
	{\cal I}_{0\perp} = \sigma_{1\perp}^{\ell_1} \sigma_{2\perp}^{\ell_2} e^{-i\bb_\perp \cdot \bP_\perp}
	\left(\frac{d}{dt_1}\right)^{\ell_1} \left(\frac{d}{dt_2}\right)^{\ell_2} 
	\left[\int d^2 k_{1\perp}\, e^{S(t_1,t_2)}\right]\Bigg|_{t_1=t_2=0}\,,\label{I0-perp-diff-form}
\end{equation}
where the function in the exponent is
\begin{equation}
	S(t_1,t_2) = t_1 (\bk_{1\perp}\cdot \bbe_+) + t_2 \left[(\bP_\perp \cdot\bbe_+) - (\bk_{1\perp} \cdot \bbe_+)\right] 
	-\frac{k_{1\perp}^2 \sigma_{1\perp}^2}{2} -\frac{(\bP_\perp - \bk_{1\perp})^2 \sigma_{2\perp}^2}{2} 
	+ i\bb_\perp \cdot \bk_{1\perp}\,.\label{function-S}
\end{equation}
This function $S(t_1,t_2)$ is a quadratic function in the components of $\bk_{1\perp}$:
\begin{equation}
	S(t_1,t_2) = -\frac{1}{2} k_{1\perp}^2 \sigma_{tot}^2 + i (\bQ\cdot \bk_{1\perp})
	+ t_2 (\bP_\perp \cdot\bbe_+) -\frac{P_\perp^2\sigma_{2\perp}^2}{2} \,,\label{function-S2}
\end{equation}
where we introduced $\sigma^2_{tot} \equiv \sigma_{1\perp}^2 + \sigma_{2\perp}^2$.
The complex valued vector $\bQ$ is
\begin{equation}
	\bQ = \bb_\perp - i \sigma_{2\perp}^2\bP_\perp  - i(t_1-t_2)\bbe_+\,,
\end{equation}
so that
\begin{equation}
	\bQ^2 = (\bb_\perp - i \sigma_{2\perp}^2\bP_\perp)^2 - 2i(t_1-t_2) (\bb_\perp - i \sigma_{2\perp}^2\bP_\perp)\cdot \bbe_+\,.
\end{equation}
Note that $\bQ^2$ is still linear in $t_1, t_2$ due to $\bbe_+ \cdot \bbe_+ = 0$.
One can then perform the standard linear shift in the $\bk_{1\perp}$ space,
\begin{equation}
	S(t_1,t_2) = -\frac{1}{2} \left(\bk_{1\perp} - \frac{i \bQ}{\sigma_{tot}^2}\right)^2 
	\sigma_{tot}^2 - \frac{1}{2}\frac{\bQ^2}{\sigma_{tot}^2}  
	+ t_2 (\bP_\perp \cdot\bbe_+) -\frac{P_\perp^2\sigma_{2\perp}^2}{2} \,,\label{function-S3}
\end{equation}
and perform the Gaussian $\bk_{1\perp}$ integration:
\begin{equation}
	\int d^2 k_{1\perp}\, \exp\left[-\frac{1}{2} \left(\bk_{1\perp} - \frac{i \bQ}{\sigma_{tot}^2}\right)^2
	\sigma_{tot}^2\right]= \frac{2\pi}{\sigma_{tot}^2}\,.
\end{equation}
The transverse integral becomes
\begin{equation}
	{\cal I}_{0\perp} = \frac{2\pi \sigma_{1\perp}^{\ell_1} \sigma_{2\perp}^{\ell_2}}{\sigma_{tot}^2} 
	\exp\left[-\frac{1}{2}P_\perp^2 \Sigma_\perp^2 - \frac{1}{2}\frac{b^2}{\sigma_{tot}^2}
	- i (\bb_\perp \cdot \bP_\perp)\frac{\sigma_{1\perp}^2}{\sigma_{tot}^2}\right] \times
	\left(\frac{d}{dt_1}\right)^{\ell_1} \left(\frac{d}{dt_2}\right)^{\ell_2} 
	\, e^{\tilde{S}(t_1,t_2)}\Bigg|_{t_1=t_2=0}\,,\label{I0-perp-diff-form-2}
\end{equation}
where $\tilde{S}(t_1,t_2)$ contains the remaining terms:
\begin{equation}
	\tilde{S}(t_1,t_2) = t_1\cdot \frac{\sigma_{2\perp}^2 (\bP_\perp\cdot \bbe_+) + i (\bb_\perp\cdot \bbe_+)}{\sigma_{tot}^2}
	+ t_2 \cdot \frac{\sigma_{1\perp}^2 (\bP_\perp\cdot \bbe_+) - i (\bb_\perp\cdot \bbe_+)}{\sigma_{tot}^2}\,.
\end{equation}
Differentiation is now elementary:
\begin{equation}
	\left(\frac{d}{dt_1}\right)^{\ell_1} \left(\frac{d}{dt_2}\right)^{\ell_2} 
	\, e^{c_1 t_1 + c_2 t_2}\Bigg|_{t_1=t_2=0} = \ c_1^{\ell_1} c_2^{\ell_2}\,,\label{diff-sequence-1}
\end{equation}
and what remains is to get a better understanding of these quantities $c_1$ and $c_2$.
It is here that the two vectors $\bA_{1\perp}, \bA_{2\perp}$ defined in Eq.~\eqref{vectors-A} 
and shown in Fig.~\ref{fig-geom} become handy:
\begin{eqnarray}
	c_1 &=& \frac{\sigma_{2\perp}^2 (\bP_\perp\cdot \bbe_+) + i (\bb_\perp\cdot \bbe_+)}{\sigma_{tot}^2}
	= \frac{\sigma_{2\perp}^2}{\sigma_{tot}^2} (\bP_\perp - \bA_{1\perp})\cdot \bbe_+ 
	= \frac{\sigma_{2\perp}^2}{\sigma_{tot}^2} |\bP_\perp - \bA_{1\perp}| e^{i\varphi_{PA1}}\,,\nonumber\\
	c_2 &=& \frac{\sigma_{1\perp}^2 (\bP_\perp\cdot \bbe_+) - i (\bb_\perp\cdot \bbe_+)}{\sigma_{tot}^2}
	= \frac{\sigma_{1\perp}^2}{\sigma_{tot}^2} (\bP_\perp - \bA_{2\perp})\cdot \bbe_+ 
	= \frac{\sigma_{1\perp}^2}{\sigma_{tot}^2} |\bP_\perp - \bA_{2\perp}| e^{i\varphi_{PA2}}\,.
\end{eqnarray}
where $\varphi_{PAi}$ is the azimuthal angle of $\bP_\perp - \bA_{i\perp}$.
Our final result is
\begin{eqnarray}
	{\cal I}_{0\perp}(\ell_1, \ell_2 > 0) &=& 
	\frac{2\pi}{\sigma_{tot}^2} \frac{(\Sigma_{\perp}^2)^{\ell_1+\ell_2}}{\sigma_{1\perp}^{\ell_1}\sigma_{2\perp}^{\ell_2}} 
	\exp\left[-\frac{1}{2}P_\perp^2 \Sigma_\perp^2 - \frac{1}{2}\frac{b^2}{\sigma_{tot}^2}
	- i (\bb_\perp \cdot \bP_\perp)\frac{\sigma_{1\perp}^2}{\sigma_{tot}^2}\right] \nonumber\\
	&& \times \left(|\bP_\perp - \bA_{1\perp}| e^{i\varphi_{PA1}}\right)^{\ell_1}\cdot
	\left(|\bP_\perp - \bA_{2\perp}| e^{i\varphi_{PA2}}\right)^{\ell_2}\,.\label{I0-prep-final-same}
\end{eqnarray}
This form makes it evident that ${\cal I}_{0\perp}$ contains two separate vortices in the $\bP_\perp$ space:
one around $\bP_\perp = \bA_{1\perp}$ with the winding number $\ell_1$ and 
the other around $\bP_\perp = \bA_{2\perp}$ with the winding number $\ell_2$.
In the limit $b\to 0$, both vectors $\bA_{i\perp} \to 0$, and the two separate 
vortices merge into a single vertex at the origin:
\begin{equation}
	{\cal I}_{0\perp}(\ell_1, \ell_2 > 0, b=0) = \frac{2\pi}{\sigma_{tot}^2} 
	\frac{(\Sigma_{\perp}^2)^{\ell_1+\ell_2}}{\sigma_{1\perp}^{\ell_1}\sigma_{2\perp}^{\ell_2}} 
	\exp\left(-\frac{P_\perp^2 \Sigma_\perp^2}{2}\right)
	(P_\perp e^{i\varphi_P})^{\ell_1+\ell_2}\,,
\end{equation}
which matches Eq.~\eqref{I0perp-zero-b}.

For the case of same-sign but negative $\ell_1, \ell_2$, we repeat the analysis using $\bbe_-$ instead of $\bbe_+$
and replace $\ell_i \to |\ell_i|$ in powers and differentiation. 
The second line of Eq.~\eqref{I0-prep-final-same} reads now
\begin{equation}
	|\bP_\perp + \bA_{1\perp}|^{|\ell_1|} e^{i\ell_1\varphi_{PA1}}\cdot
	|\bP_\perp + \bA_{2\perp}|^{|\ell_2|} e^{i\ell_2\varphi_{PA2}}\,,
\end{equation}
with the same vectors $\bA_{i\perp}$ as before and with $\varphi_{PAi}$ being now the azimuthal angles
of $\bP_\perp + \bA_{i\perp}$.

\subsection{Non-zero impact parameter: $b \not = 0$, opposite-sign $\ell_1, \ell_2$}

If $\ell_1$ and $\ell_2$ are of the opposite signs, the above procedure needs to be slightly modified.
For definiteness, support $\ell_1 < 0$ and $\ell_2 > 0$. The expression \eqref{I0-perp-diff-form}
still holds, with $\ell_1 \to |\ell_1|$ but the exponent function $S(t_1,t_2)$ 
now differs from Eq.~\eqref{function-S} by using $\bbe_-$ instead of $\bbe_+$ in the $t_1$ term.
This bears consequences for the shift vector $\bQ$, which is now
\begin{equation}
	\bQ = \bb_\perp - i \sigma_{2\perp}^2\bP_\perp  - i t_1 \bbe_- + i t_2\bbe_+\,,
\end{equation}
so that its square
\begin{equation}
	\bQ^2 = (\bb_\perp - i \sigma_{2\perp}^2\bP_\perp)^2 
	- 2i t_1 (\bb_\perp - i \sigma_{2\perp}^2\bP_\perp)\cdot \bbe_-
	+ 2i t_2 (\bb_\perp - i \sigma_{2\perp}^2\bP_\perp)\cdot \bbe_+ + 4 t_1 t_2
\end{equation}
now contains a new term $2 t_1 t_2$.
The Gaussian integral remains as before, the expression in Eq.~\eqref{I0-perp-diff-form-2}
still holds, with $\ell_1 \to |\ell_1|$, the only difference being that 
$\tilde S(t_1,t_2)$ features now $\bbe_-$ in the $t_1$ term and contains an extra piece 
$-2t_1t_2/\sigma_{tot}^2 \equiv -c_{12} t_1 t_2$.

This new term may look dangerous but there is an elegant way of dealing with it.
We need to work out the following differentiation:
\begin{equation}
	\left(\frac{d}{dt_1}\right)^{|\ell_1|} \left(\frac{d}{dt_2}\right)^{\ell_2} 
	\, e^{c_1 t_1 + c_2 t_2 - c_{12} t_1t_2}\Bigg|_{t_1=t_2=0}\,.\label{diff-sequence-2}
\end{equation}
We first shift the variables $t_i$:
\begin{equation}
	c_1 t_1 + c_2 t_2 - c_{12} t_1t_2 = -c_{12}\left(t_1 - \frac{c_2}{c_{12}}\right) \left(t_2 - \frac{c_1}{c_{12}}\right) + \frac{c_1c_2}{c_{12}}
	\equiv -c_{12} \bar t_1 \bar t_2  + \frac{c_1c_2}{c_{12}}\,.
\end{equation}
The derivatives can now be done with respect to $\bar t_1, \bar t_2$ evaluated at the shifted points
$\bar t_1 = -c_2/c_{12}$ and $\bar t_2 = -c_1/c_{12}$.
Then we check which of the derivatives is of a higher order and take them first. For example, suppose $\ell_2 \ge |\ell_1|$. Then 
\begin{equation}
	e^{c_1c_2/c_{12}}\left(\frac{d}{d\bar t_1}\right)^{|\ell_1|} \left(\frac{d}{d\bar t_2}\right)^{\ell_2} 
	\, e^{-c_{12} \bar t_1\bar t_2}\Bigg|_{\raisebox{3mm}{\scriptsize$\begin{array}{l}\bar t_1=-c_2/c_{12}\\ \bar t_2=-c_1/c_{12} \end{array}$}}
	\ = \ (-c_{12})^{\ell_2} \,e^{c_1c_2/c_{12}}\left(\frac{d}{d\bar t_1}\right)^{|\ell_1|} \left[(\bar t_1)^{\ell_2} e^{c_1 \bar t_1}\right]\Bigg|_{\bar t_1=-c_2/c_{12}}	\,.\label{diff-sequence-3}
\end{equation}
We now write $\ell_2 = |\ell_1| + k$, with $k \ge 0$, and apply the Rodrigues formula for the associated Laguerre polynomials
\begin{equation}
	L_n^k(x) = \frac{x^{-k}e^x}{n!}\left(\frac{d}{dx}\right)^n\left(x^{n+k}e^{-x}\right)
\end{equation}
to obtain
\begin{equation}
	(-c_{12})^{|\ell_1|}\, |\ell_1|!\, c_2^{\ell_2 - |\ell_1|}\, 
	L_{|\ell_1|}^{\ell_2 - |\ell_1|}\left(\frac{c_1 c_2}{c_{12}}\right)\,.
\end{equation}
Note that for this arrangements of $\ell_1$ and $\ell_2$, we can write $\ell_2 - |\ell_1| = \ell_1 + \ell_2$
and $|\ell_1| = \ell_-$. Thus, the final expression for the transverse integral is
\begin{eqnarray}
	{\cal I}_{0\perp}(\ell_1 < 0, \ell_2 \ge |\ell_1|) &=& 
	(-2)^{\ell_-}\, \ell_{-}!\,\frac{2\pi}{\sigma_{tot}^2} 
	\frac{(\Sigma_\perp^2)^{|\ell_1+\ell_2|+\ell_-}}{\sigma_{1\perp}^{|\ell_1|} \sigma_{2\perp}^{\ell_2}}	
	\exp\left[-\frac{1}{2}P_\perp^2 \Sigma_\perp^2 - \frac{1}{2}\frac{b^2}{\sigma_{tot}^2}
	- i (\bb_\perp \cdot \bP_\perp)\frac{\sigma_{1\perp}^2}{\sigma_{tot}^2}\right] \nonumber\\[2mm]
	&& \hspace{-2cm}\times\ 	|\bP_\perp - \bA_{2\perp}|^{|\ell_1+\ell_2|} e^{i(\ell_1+\ell_2)\varphi_{PA2}}
	\cdot L_{\ell_-}^{|\ell_1+\ell_2|}\left(\frac{\Sigma_\perp^2}{2}|\bP_\perp + \bA_{1\perp}| e^{-i\varphi_{PA1}}
	\cdot |\bP_\perp - \bA_{2\perp}| e^{i\varphi_{PA2}}\right)
\,.\label{I0-prep-final-opposite}
\end{eqnarray}
The same calculation for $|\ell_1| > \ell_2 = \ell_-$ gives the identical result apart from the position of 
vortex and contains the factor ${|\bP_\perp + \bA_{1\perp}|}^{|\ell_1+\ell_2|} e^{i(\ell_1+\ell_2)\varphi_{PA1}}$ in the second line.
In the limit $b \to 0$, these results coincide with Eq.~\eqref{I0perp-zero-b}. 
Alternatively, if $\ell_- = 0$, then Eq.~\eqref{I0-prep-final-opposite} coincides with Eq.~\eqref{I0-prep-final-same}
taken under the same assumption $\ell_1 = 0$.


\end{document}